\documentclass{aa}
\usepackage{graphics}
\newcommand{\mm}{\mu\mathrm{m}}
\newcommand{\lambdaJ}{\lambda_{\mathrm{J}}}
\newcommand{\lambdaH}{\lambda_{\mathrm{H}}}
\newcommand{\lambdaK}{\lambda_{\mathrm{K}}}
\newcommand{\logd}{\log_{10}}
\newcommand{\Ae}{A_{\mathrm{e}}}
\newcommand{\Tp}{T_{\mathrm{p}}}
\newcommand{\BT}{B_{\mathrm{T}}}
\newcommand{\Rvc}{R_{25}}
\newcommand{\Dvc}{D_{25}}

\newcommand{\NIR}{\mathop{\mathrm{NIR}}}
\begin{document}
\thesaurus{03(11.05.2; 11.06.2; 11.16.1; 11.19.7; 13.09.1; 09.04.1)}
\title{A statistical study of nearby galaxies}
\subtitle{I: NIR growth curves and optical-to-NIR colors
as a function of type, luminosity and inclination}
\author{Michel Fioc\inst{1,2} \and Brigitte
Rocca-Volmerange\inst{2,3}}
\institute{NASA/Goddard Space Flight Center, code 685, Greenbelt, MD 20771, USA
\and Institut d'Astrophysique de Paris, CNRS,
98 bis Bd Arago, F-75014 Paris, France 
\and Institut d'Astrophysique Spatiale,
B\^at. 121, Universit\'e Paris XI, F-91405 Orsay, France}
\offprints{Michel Fioc}
\mail{fioc@zardoz.gsfc.nasa.gov}
\date{Received / Accepted}
\titlerunning{Optical-to-NIR colors of nearby galaxies}
\authorrunning{M. Fioc \& B. Rocca-Volmerange}
\maketitle
\begin{abstract}
Growth curves of the near-infrared (NIR) magnitude 
as a function of the aperture
have been built and used to derive NIR total magnitudes
from aperture data taken from the literature.
By cross-correlating with optical and redshift data, absolute 
magnitudes and optical-to-NIR colors have been computed 
for some 1000 galaxies of different types. Significant
color gradients are observed, underlining that small aperture
colors may lead to a biased picture of the stellar populations
of galaxies.

A statistical analysis using various estimators taking into account
the intrinsic scatter has been performed to establish
relations between the colors, the morphological type,
the inclination or the shape, and the intrinsic luminosity.

The combination of the optical and the NIR should obviously improve 
our understanding of the evolution of galaxies.
Despite the intrinsic scatter, especially among star-forming galaxies,
optical-to-NIR colors show a very well defined sequence with type,
blueing by 1.3 mag from ellipticals to irregulars.
The colors of spiral galaxies strongly redden with increasing 
inclination and put new constraints on the modeling of the extinction.
No such effect is observed for lenticular galaxies.
We also find that rounder ellipticals tend to be redder.

A color-absolute magnitude relation is observed inside each type,
with a slope significantly steeper for early and intermediate
spirals than for ellipticals or late spirals.
This stresses the importance of considering both the mass and the type
to study the star formation history of galaxies.
\end{abstract}
\section{Introduction}

Although color-magnitude diagrams have now been obtained for a few 
very close galaxies -- most of them dwarf --, allowing to study 
directly their stellar populations (e.g. Schulte-Ladbeck \& Hopp
\cite{SLH}; Aparicio \cite{Aparicio}), the synthesis of the spectral
energy distribution (SED) of nearby galaxies remains the most
efficient and systematic method to trace their star formation history
up to $z=0$.

Early optical studies have shown a clear correlation between the
SED -- and hence the star formation history -- and the galaxy Hubble type.
However, the age-star formation timescale and age-metallicity 
degeneracies make it highly desirable to extend the wavelength range
to the ultraviolet (UV) and the near-infrared. Whereas the ultraviolet
is related to the young stellar populations in currently star-forming galaxies,
the near-infrared is dominated by old giant stars and is a measure of the star
formation rate integrated from the beginning, i.e. of the stellar
mass of the galaxy. When compared to shorter wavelengths, the NIR
may moreover put constraints on the stellar metallicity, owing to the
high sensitivity of the effective temperature of red giants
to the abundance of heavy elements.
In combination with optical data, it finally provides clues on the amount and the
distribution of dust because of the different extinction at both wavelengths.

Observations in the UV and the NIR are however hampered by the atmosphere.
Though interesting to study the 
detailed star formation history of a specific galaxy, the spectra
obtained from Space are too scarce to afford a statistical analysis 
from which the general trends as a function of type or mass, or the
effect of the dust, may be derived.
In spite of the loss of spectral resolution,
broad-band colors in the $J$, $H$ and $K$ atmospheric windows 
are more promising.
As the spectra, they are usually obtained in small apertures
not representative of the whole galaxy
because of the color gradients, especially for disk galaxies, 
and have to be extrapolated to be compared to the optical data.
Such extrapolation is however much easier for colors than for spectra
and can be applied to hundreds of objects.

In the following, we analyze a catalog of infrared aperture magnitudes
(section~\ref{catalogue}) and combine it with optical catalogs to
derive NIR growth curves of the magnitude as a function of the
aperture (section~\ref{croissance}).
We then compute total and effective magnitudes and colors (section~\ref{magnitudes}). 
The NIR and optical-to-NIR colors are analyzed statistically as a function of
type, luminosity and inclination in section~\ref{statistiques}. We finally discuss
our results in section~\ref{discussion}.
\section{The optical and infrared data}
\label{catalogue}
The near-infrared data used in this paper come from the CIO catalog
(Gezari et al. \cite{Gezari}) which is a compilation 
of the NIR observations published before 1995. 
They are given in a large variety of systems (magnitudes $m_{\lambda}$,
$F_{\lambda}$, $F_{\nu}$, etc.)
and units, and are
first converted to magnitudes assuming Vega has a magnitude of 0.03.
The data are
reduced from the wavelength $\lambda$ 
to the central wavelengths of the $J$ ($\lambdaJ=1.24\mm$),
$H$ ($\lambdaH=1.65\mm$) and $K$ ($\lambdaK=2.21\mm$) filters of
Bessel \& Brett (\cite{BB}), assuming the following color equations:
\[
m_{\mathrm{J}}=m_{\lambda}+(\lambdaJ-\lambda)\frac{<J-H>}{\lambdaJ-\lambdaH},\]
\begin{eqnarray*}
m_{\mathrm{H}}&=&m_{\lambda}+(\lambdaH-\lambda)\frac{<J-H>}{\lambdaJ-\lambdaH}
\mathrm{\qquad{}if\;}\lambda<1.65\mm,\\
&=&m_{\lambda}+(\lambdaH-\lambda)\frac{<H-K>}{\lambdaH-\lambdaK}
\mathrm{\qquad{}if\;}\lambda>1.65\mm,
\end{eqnarray*}
\[m_{\mathrm{K}}=m_{\lambda}+(\lambdaK-\lambda)\frac{<H-K>}{\lambdaH-\lambdaK},\]
where $<J-H>=0.7$ and $<H-K>=0.2$ are the typical $J-H$ and $H-K$
colors of a normal galaxy and depend very little on the galaxy type
or the aperture.

The CIO catalog is cross-correlated with the RC3 catalog (de
Vaucouleurs et al. \cite{RC3}) to
correct for the Galactic extinction. The NIR extinction is computed
from $A_{\mathrm{B}}$ as  follows: $A_{\mathrm{J}}=0.182 A_{\mathrm{B}},\;A_{\mathrm{H}}=0.098 A_{\mathrm{B}},\;A_{\mathrm{K}}=0.068 A_{\mathrm{B}}$.

The magnitudes are also corrected for the redshift $z$
as $m_{\lambda}(0)=m_{\lambda}(z)-k_{\lambda}(z)$. 
Values of $k_{\lambda}(z)$ are given in Table~\ref{corrk}
for different types of galaxies
and have been computed using the \textsc{p\'egase} model of
spectral evolution (Fioc \& Rocca-Volmerange \cite{FRV}).
They should be used only at $z<0.1$.
The redshift is taken from the NED database. 
For a few galaxies, $z$ is unknown and we compute the $k$-correction
(but not the absolute magnitudes)
assuming $z=0.01$, the mean redshift of our catalog.
\begin{table}
\begin{center}
\begin{tabular}{l|c|c|c|c}
type&$k_{\mathrm{B}}$    & $k_{\mathrm{J}}$    &$k_{\mathrm{H}}$    &$k_{\mathrm{K}}$    \\
\hline
E&$5.1z$   & $-0.3z$  &$-0.1z$  &$-2.8z$  \\
S0&$4.9z$   & $-0.3z$  &$-0.1z$  &$-2.8z$  \\
Sa&$4.0z$   & $-0.3z$  &$-0.1z$  &$-2.8z$  \\
Sb&$3.6z$   & $-0.3z$  &$-0.1z$  &$-2.8z$  \\
Sc&$2.6z$   & $-0.3z$  &$-0.1z$  &$-2.8z$  \\
Sd&$2.2z$   & $-0.5z$  &$-0.3z$  &$-2.8z$  \\
Im&$1.8z$   & $-0.7z$  &$-0.4z$  &$-2.8z$   
\end{tabular}
\caption{$k$-corrections in $B$, $J$, $H$ and $K$ as a function of the
type.}
\label{corrk}
\end{center}
\end{table}

The RC3 provides us also with the morphological type $T$, 
 the ratio $\Rvc$ of the major axis ($\Dvc$) to the minor axis 
of the ellipsis corresponding to the isophot $\mu=25\mathrm{mag/arcsec}^2$,
the total $B$-magnitude $\BT$ extrapolated to an infinite radius
and the circular effective aperture $\Ae$
containing half the light emitted in the $B$-band.

When available, we prefer to take $\BT$ and $\Ae$
from the more recent catalog Hypercat (Prugniel \& H\'eraudeau
\cite{PH}). 
This catalog
also gives the photometric type $\Tp$ corresponding to the best-fitting
growth curve ${\cal B}(X,\Tp)$ of the $B$-magnitude as a function of
the circular aperture $A$, where $X=\logd(A/\Ae)$ (see Appendix~A).
\section{The NIR growth curve}
\label{croissance}
The detailed 2D-fitting of the profile of galaxies (de Jong 
\& van der Kruit \cite{dJvdK})
is certainly the best way to compute their total magnitudes, but 
surface photometry in the NIR is available for too few
of them to perform a statistical analysis. The largest such study 
is based on only 86 spiral galaxies of all types 
(de Jong \cite{dJa,dJb,dJc}), to compare to
typically 100 galaxies {\em per type} in this paper.
We therefore prefer
to extrapolate aperture magnitudes with a growth curve.
The computation of total (i.e. asymptotic)
NIR magnitudes comparable to the optical
ones determined by, e.g., de Vaucouleurs et al. (\cite{RC3})
or Prugniel \& H\'eraudeau (\cite{PH}) is however made difficult by the
small apertures achieved at these wavelengths. Few galaxies
have enough data from small to large apertures to constrain
the shape of the growth curve and one has to use another method,
combining observations of different galaxies having presumably
the same curve (cf. Griersmith \cite{Griersmith}).
To this purpose, one needs to scale the apertures to a 
characteristic length $A_{\mathrm{c}}$ for each galaxy. The NIR growth
curve 
${\cal M}(A/A_{\mathrm{c}})$
is then built by plotting $m(A)-m(A_{\mathrm{c}})$ as a function of
$\logd(A/A_{\mathrm{c}})$.
Usually (Gavazzi et al. \cite{Gavazzi}; Gavazzi \& Boselli \cite{GB};
Tormen \& Burstein \cite{TB}; Frogel et al. \cite{Frogel}; Aaronson et
al. \cite{Aaronson}), the $\Dvc$ 
diameter has been used as characteristic length. As noticed
however by de Vaucouleurs et al. (\cite{RC2}), the $\Dvc$ diameter depends not only 
on the shape of the profile but also on the central surface
brightness. For this reason, the ratio of the effective aperture $\Ae$ to $\Dvc$ 
is not a constant. 

We prefer to adopt here an other procedure. We may expect some
relation between the optical profile and the NIR profile.
For this reason, we decide to build a growth curve as a function
of the $B$ {\em photometric} type $\Tp$ taken from Hypercat and of 
$X=\logd(A/\Ae)$. The infrared magnitude $m(\Ae)$ in the $B$ 
effective aperture is computed by interpolation 
or by a small extrapolation in $\logd(A)$.
Rather than a straight line,
we have used the functions proposed to compute 
$H$ magnitudes at $A=\Dvc$ by Gavazzi et al. (\cite{Gavazzi}):
\[{\cal
M}(X_{25})=a+b(-0.35X_{25}+1.68X_{25}^2+0.07X_{25}^3)\] 
where $X_{25}=\logd(A/\Dvc)$,
and fitted $a$ and $b$ to the observations.
This takes into account the curvature of the growth curve
near $\Ae$ and improves the determination of $m(\Ae)$.
Because they are polynomials, these functions are however not suitable 
to extrapolate the magnitude to the infinity.

The $J$, $H$ and $K$ data have been combined to build the growth curve.
This is justified by the fact that only small $J-H$ and $H-K$ color gradients 
are observed (Aaronson \cite{Aaronsonthese}; Frogel et al. \cite{Frogel}).
\begin{figure*}
\begin{center}
\resizebox{!}{21cm}{\rotatebox{0}{\includegraphics{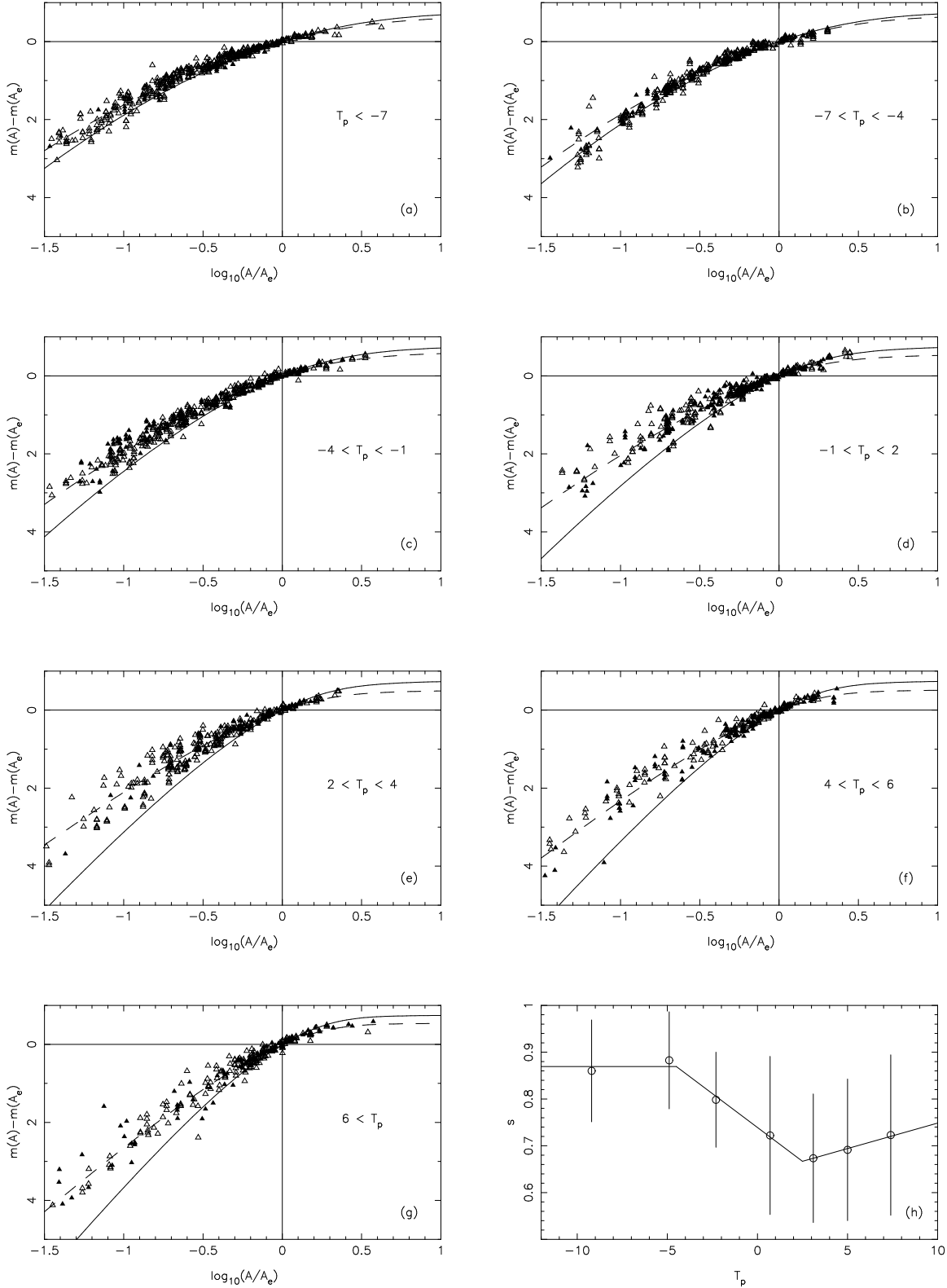}}}
\caption{{\bf (a) to (g):} Growth curve of the NIR magnitude as a
function of the aperture for the different photometric types.
Hollow and filled triangles correspond respectively 
to galaxies with $\logd R_{25}<0.3$ and
$\logd R_{25}>0.3$. The solid and dashed lines are the mean optical and
NIR growth curves. They are
normalized to the effective aperture in the $B$-band and to the 
corresponding magnitude.
{\bf (h):} The circles and vertical segments are the mean value $s_0$
and the intrinsic scatter $\sigma_{s_0}$ of $s$ as a function of the
median value of $\Tp$ in each bin. The broken line
is the 3-slope fit to $s_0(\Tp)$ given in the text.}
\label{magouv}
\end{center}
\end{figure*}
The growth curves are plotted in Fig.~\ref{magouv} for 7 bins of
photometric type. As evidenced by this graph, the NIR growth curve
is flatter than the optical one, especially for $\Tp$ corresponding 
to intermediate spirals: the NIR emission of galaxies is 
more concentrated than in the optical.
This behavior is expected because the bulge is redder than the
disk and is therefore more prominent in the NIR than in the optical.
The scatter is also higher for intermediate types, which
might be due to an inclination dependency of the growth curve (Christensen
\cite{Christensen}; Kodaira et al. \cite{Kodaira}). We have therefore
distinguished in Fig.~\ref{magouv} between galaxies with $\Rvc<0.3$ 
(rather face-on for disk galaxies) and $\Rvc>0.3$, but no obvious trend
is observed and we will neglect any inclination dependency in the following.

Because of the scatter and the lack of data at high aperture,
where the flattening of the growth curve puts constraints 
on the curvature, 
attempts to fit growth curves similar to those of 
Prugniel \& H\'eraudeau (\cite{PH}) with 3 parameters
($m_{\mathrm{T}}$, $\logd(\Ae)$ and $\Tp$) 
did not prove successful.
Plotting the NIR magnitudes versus the $B$ magnitudes
however reveals a nearly linear -- though scattered -- 
relation, between these quantities, suggesting 
to adopt a NIR growth curve of this form:
\[{\cal M}(X,\Tp)=s{\cal B}(X,\Tp).\]
This ensures that the extrapolation at infinite $A$
converges since
$\lim_{X\rightarrow\infty}{\cal B}(X,\Tp)=0$.
Such relation has been fitted for each photometric type bin (Fig.~\ref{magouv},
dashed lines) at $X>-1.5$.

The dispersion is due both to the uncertainties in the
individual data and to the intrinsic scatter in the shape of the 
NIR growth curves for a given $B$ photometric type.
The following convention is adopted hereafter:
{\em $x \sim {\cal N}(\mu,\sigma^2)$} means that {\em $x$ is distributed
according to (or the density probability of $x$ is) a Gaussian 
with mean $\mu$ and variance $\sigma^2$}.
Let us assume that, for each galaxy, the distribution of individual
data around the best-fitting NIR growth curve 
$s{\cal B}(X,\Tp)$ is $\sim {\cal N}(0,\sigma^2_{m_0})$
and that the distribution of the parameter $s$ characterizing the
growth curve is $s\sim {\cal N}(s_0,\sigma^2_{s_0})$.
At any aperture, 
\begin{eqnarray*}
m(X)-m(0)\sim {\cal N}\Bigl\lbrace\!\!&\!\!&\!\!s_0\bigl[{\cal B}(X,\Tp)-{\cal B}(0,\Tp)\bigr],\\
\!\!&\!\!&\!\! \sigma_{m_0}^2+\sigma_{s_0}^2\bigl[{\cal B}(X,\Tp)-{\cal
B}(0,\Tp)\bigr]^2
\Bigr\rbrace.
\end{eqnarray*}
A maximum likelihood estimation of these parameters yields
$\sigma_{m_0}\simeq 0.06$, nearly independent of $\Tp$, 
$\sigma_{s_0}\simeq 0.10$ for early types ($\Tp < -1$) and
$\sigma_{s_0}\simeq 0.15$ at later types.
The value of $s_0$ we obtained is plotted as a
function of the median $\Tp$ of each bin in Fig.~\ref{magouv}
and may be approximated with the following formulae:
\begin{eqnarray*}
s_0&=&0.87\mathrm{\qquad{}if\;}\Tp<-4.5,\\
&=&0.74-0.029\Tp\mathrm{\qquad{}if\;}-4.5<\Tp<2.5,\\
&=&0.64+0.011\Tp\mathrm{\qquad{}if\;}\Tp>2.5.
\end{eqnarray*}
The mean value $s_0$ is less than one for all types, corresponding
as expected to a blue-outwards gradient. This gradient is low for 
early-types, increases for spirals, peaking at $\Tp\in [2-4]$ ($\sim$~Sb) 
in good agreement with what has been obtained at optical wavelengths
by de Vaucouleurs \& Corwin (\cite{dVC}) and remains constant or
slightly decreases for late types. 
The typical blueing of the $B-\NIR$ color of spirals from the
effective aperture to the infinity is $-0.75(1-s)\sim -0.2$, 
close to the median value ($-0.19$) in $B-H$ and $B-K$
we have computed from the profiles published by de Jong (\cite{dJa}).
\section{Effective and total near-infrared magnitudes}
\label{magnitudes}
\subsection{Apparent magnitudes and colors}
Effective and total near-infrared magnitudes have been computed
for all the galaxies with known effective aperture by fitting the growth curves $s{\cal B}(X,\Tp)$ 
introduced in the previous section to
the observed data $(X_i, m_i, \sigma_i),\;i\in[1,n]$, where 
$\sigma_i$ is the uncertainty on the magnitude $m_i$. 
The asymptotic magnitude depends primarily on the shape of the growth curve
at large aperture. To avoid problems with small apertures suffering
from seeing or off-centering, or which are contaminated by nuclear
emission or a central starburst, we remove all the observations
at $X<-1$ from the fit.

The procedure to compute the
total and effective NIR magnitudes $m_{\mathrm{T}}$, $m_{\mathrm{e}}$ 
and the corresponding
uncertainties $\sigma(m_{\mathrm{T}})$, $\sigma(m_{\mathrm{e}})$ 
due to the uncertainties in the aperture magnitudes 
is detailed in Appendix~B.

The global uncertainty $\varsigma_{\mathrm{m_{\mathrm{T}}}}$ on 
$m_{\mathrm{T}}$ taking also into account the uncertainties 
in $\Tp$ and $\Ae$ is finally computed from 
\[\varsigma_{m_{\mathrm{T}}}^2=\sigma^2(m_{\mathrm{T}})+\left(\frac{\partial
m_{\mathrm{T}}}{\partial \Tp}\right)^2\sigma^2(\Tp)+\left(\frac{\partial
m_{\mathrm{T}}}{\partial \Ae}\right)^2\sigma^2(\Ae)\]
and the same for $\varsigma_{m_{\mathrm{e}}}$.

In many cases, the photometric type
is unknown and we estimate it from the morphological type
with the following relations derived by cross-correlating
Hypercat and the RC3:
\begin{eqnarray*}
\Tp&=&-2.3+0.5T\mathrm{\qquad{}if\;}T<-1,\\
&=&-1.32+1.48T\mathrm{\qquad{}if\;}-1<T<4,\\
&=&1.4+0.8T\mathrm{\qquad{}if\;}T>4.
\end{eqnarray*}
We then simply assume $\sigma(\Tp)=3$.

The uncertainty on the colors $m_1-m_2$, where $(m_1,m_2)\in [J,H,K]^2$,
is computed as 
\begin{eqnarray*}\varsigma^2_{m_1-m_2}=\sigma^2(m_1)+\sigma^2(m_2)
\!&\!+\!&\!\left[\frac{\partial
(m_1-m_2)}{\partial \Tp}\right]^2\sigma^2(\Tp)\\\!&\!+\!&
\!\left[\frac{\partial
(m_1-m_2)}{\partial \Ae}\right]^2\sigma^2(\Ae).\end{eqnarray*}
However, this is clearly an overestimate because $J$, $H$ and $K$ data
have usually been observed simultaneously and their errors
are presumably highly positively correlated.

When $B$ is one of the band, e.g. $m_1=B_{\mathrm{T}}$, 
the partial derivatives are unknown.
We then estimate $\varsigma_{m_1-m_2}$ as
\[\varsigma^2_{m_1-m_2}=\varsigma^2_{m_1}+\varsigma^2_{m_2}.\]
This will usually be a slight overestimate for the partial 
derivatives of $\BT$ and $m_2$ should have the same sign, but most 
of the uncertainty comes actually from the NIR data themselves.
\subsection{Absolute magnitudes}
The absolute magnitude $M_{\mathrm{T}}$ is computed as
\[M_{\mathrm{T}}=m_{\mathrm{T}}-25-5\logd(v_{\mathrm{V}}/H_0)\]
where $v_{\mathrm{V}}$ is the velocity of a galaxy derived 
from the redshift $z$ and corrected for the movement of the
Sun in the restframe of the Virgo cluster according to the equations
(18), (19), (20), (31) and (32) from Paturel 
et al. (\cite{Paturel}).
We assume $H_0=65\,\mathrm{km.s^{-1}.Mpc^{-1}}$.
The uncertainty on the absolute magnitude is computed 
from 
\[\varsigma^2_{M_{\mathrm{T}}}=\varsigma^2_{m_{\mathrm{T}}}+\left[\frac{5}{\ln(10)}\right]^2\frac{(\partial
v_{\mathrm{V}}/\partial z)^2\sigma^2(z)+v^2_{\mathrm{p}}}{v^2_{\mathrm{V}}}\]
where $v_{\mathrm{p}}=350\,\mathrm{km.s^{-1}}$ is the typical peculiar 
velocity of galaxies in the Las Campanas Redshift Survey (Lin et
al. \cite{LCRS}). At low redshift, peculiar velocities perturb
the redshift-distance relation and increase the uncertainty on
the absolute magnitude.
All the absolute magnitudes of the galaxies with $[(\partial
v_{\mathrm{V}}/\partial
z)^2\sigma^2(z)+v^2_{\mathrm{p}}]^{1/2}>v_{\mathrm{V}}/3$ have been 
discarded in the following.
$\sigma(z)$ has been computed according to the formulae (1) and (2) of
the Appendix A of Paturel et al. (\cite{Paturel}), taking into account
both the internal uncertainty on the values of the redshift and the
discrepancies between them.

Covariances between the colors and the absolute magnitudes have also
been computed.
\section{Statistics of optical-near infrared colors}
\label{statistiques}
\subsection{$J-H$, $H-K$ and $J-K$ colors}
\begin{figure*}
\resizebox{!}{22cm}{\rotatebox{0}{\includegraphics{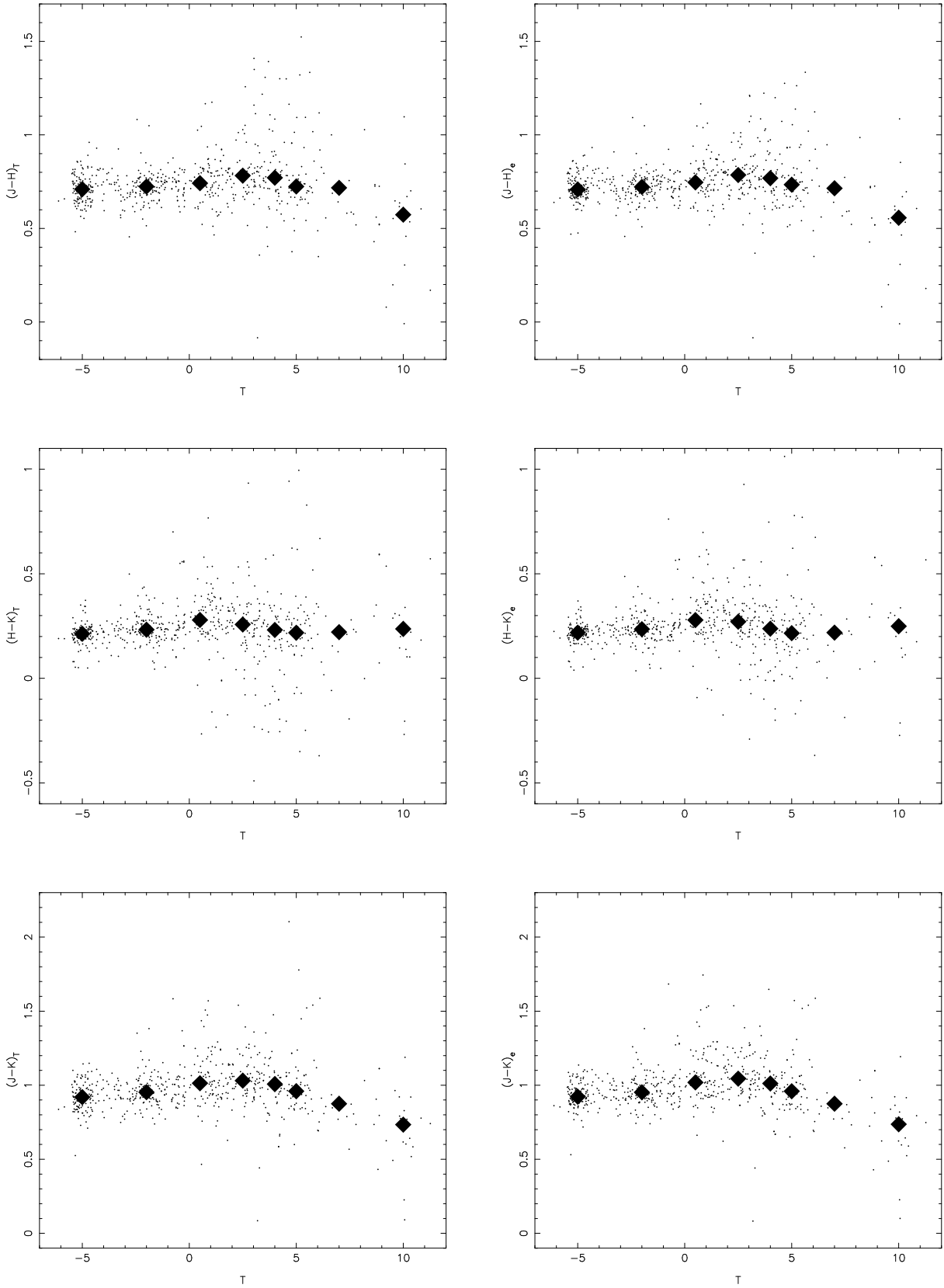}}}
\caption{Total and effective NIR colors as a function of type.
A random component in $[-0.5, 0.5]$ has been added to the type 
to make the graph clearer. The diamonds are the median colors per type.}
\label{figTJHK}
\end{figure*}

The galaxies have been distributed in eight broad types 
according to the morphological type $T$ taken from the RC3 
(see Table~\ref{types}).
\begin{table}
\begin{center}
\begin{tabular}{l|r@{}l}
type & $T\;$ & (RC3)\\
\hline
E &&$T<-3.5$\\		
S0 &$-3.5<\;$&$T<-0.5$\\	
Sa &$-0.5<\;$&$T<1.5$\\	
Sb &$1.5<\;$&$T<3.5$\\	
Sbc &$3.5<\;$&$T<4.5$\\	
Sc &$4.5<\;$&$T<5.5$\\	
Sd &$5.5<\;$&$T<8.5$\\	
Im &$8.5<\;$&$T$
\end{tabular}
\caption{Correspondence between the types used in this paper and the
RC3 types.}
\label{types}
\end{center}
\end{table}
The $J-H$, $H-K$ and $J-K$ total and effective colors have been
plotted 
in figure~\ref{figTJHK} 
as a function of the type. 

Because the uncertainties computed for $J-H$, $H-K$ and $J-K$ are
overestimates, we simply give in table~\ref{tabTJHK} the median color for
each type and compute its uncertainty by a bootstrap with
replacement.
The total and effective colors show almost no difference.
Whereas $H-K$ seems to be nearly
independent of the type, $J-H$ and $J-K$ decrease clearly at the
late types, indicating that the $J$-band is contaminated by young
stars. The colors are also redder for early-type spirals, which is 
difficult to understand with current models of stellar evolution and 
atmospheres. This effect is observed in both the total and effective
colors. The extrapolation to compute the latter being small, this may
not come from a problem with the growth curve. The extinction,
negligible in the NIR, is also unlikely to cause this phenomenon. A 
possible explanation is that we begin to see in $H$ and $K$ the tail
of the dust infrared emission at short wavelengths. We note in particular that 
the $J$ and $K$ samples contain a non-negligible fraction of galaxies
with nuclear activity of which central colors are in some cases as red
as $J-H\sim H-K\sim 1$.
\begin{table*}
\begin{center}
\begin{tabular}{l||r|c|c||r|c|c||r|c|c}
type&$N$&$(J-H)_{\mathrm{T}}$&$(J-H)_{\mathrm{e}}$&$N$&$(H-K)_{\mathrm{T}}$&$(H-K)_{\mathrm{e}}$&$N$&$(J-K)_{\mathrm{T}}$&$(J-K)_{\mathrm{e}}$\\
&&median&median&&median&median&&median&median\\
\hline
E&140&$  0.71\pm  0.01$&$  0.71\pm  0.01$&141&$  0.21\pm  0.01$&$  0.22\pm  0.01$&140&$  0.92\pm  0.01$&$  0.92\pm  0.01$\\
S0&154&$  0.72\pm  0.01$&$  0.72\pm  0.01$&157&$  0.23\pm  0.01$&$  0.23\pm  0.01$&155&$  0.95\pm  0.01$&$  0.95\pm  0.01$\\
Sa& 96&$  0.74\pm  0.01$&$  0.75\pm  0.01$& 96&$  0.28\pm  0.01$&$  0.28\pm  0.01$& 97&$  1.01\pm  0.01$&$  1.02\pm  0.01$\\
Sb& 93&$  0.78\pm  0.01$&$  0.79\pm  0.01$& 95&$  0.26\pm  0.01$&$  0.27\pm  0.01$& 95&$  1.03\pm  0.01$&$  1.04\pm  0.01$\\
Sbc& 46&$  0.77\pm  0.02$&$  0.77\pm  0.02$& 47&$  0.23\pm  0.02$&$  0.24\pm  0.02$& 47&$  1.01\pm  0.02$&$  1.01\pm  0.01$\\
Sc& 46&$  0.72\pm  0.02$&$  0.73\pm  0.02$& 46&$  0.22\pm  0.02$&$  0.22\pm  0.02$& 46&$  0.96\pm  0.02$&$  0.96\pm  0.02$\\
Sd& 26&$  0.72\pm  0.04$&$  0.71\pm  0.04$& 24&$  0.22\pm  0.04$&$  0.22\pm  0.04$& 24&$  0.87\pm  0.05$&$  0.88\pm  0.04$\\
Im& 22&$  0.57\pm  0.05$&$  0.56\pm  0.03$& 20&$  0.24\pm  0.05$&$  0.25\pm  0.04$& 25&$  0.73\pm  0.05$&$  0.74\pm  0.06$\\
\end{tabular}
\caption{Median total and effective $J-H$, $H-K$ and $J-K$ colors
(with their $1\sigma$-uncertainties) as a function
of type.}
\label{tabTJHK}
\end{center}
\end{table*}
\subsection{$B-J$, $B-H$ and $B-K$ colors}
The catalog is large enough to perform a statistical analysis
of the colors as a function of the type, the luminosity and the
inclination or the shape ($\Rvc$) of the galaxy. An interesting
quantity is also the intrinsic scatter in the colors at a given type,
luminosity and $\Rvc$, for it is a measure of the variations of the
star formation history and of the effects of dust (e.g. Peletier \&
de Grijs \cite{PdG}; Shioya \& Bekki \cite{SB}).
Note however that the intrinsic scatter may depend on the
type-binning, especially if the colors evolve rapidly from type to type.

The $B-H$ sample is by far the largest (more than 900 galaxies in $B-H$ against
600 in $B-J$ and $B-K$). It is also the most accurate
(the uncertainty is typically 0.16 mag in $(B-H)_{\mathrm{T}}$ and 0.09 mag in $(B-H)_{\mathrm{e}}$)
and the most complete in type, whereas the $B-J$ and $B-K$ sample are strongly
deficient in the latest types. For these various reasons, we will
mainly focus our discussion in the following on the analysis
of the $B-H$ data.

We adopt here the formalism proposed by Akritas (\cite{Akritas}).
Let us assume that the true color $Y_i$ of the $i^{\mathrm{th}}$ galaxy depends
linearly on $p$ quantities $X_{ij}, j\in[1,p]$ (the $X_{ij}$ being, e.g., 
$\logd(\Rvc)$ or the absolute magnitude):
\[Y_i=\beta_0+\sum_{j=1}^p\beta_j(X_{ij}-\mu_{j})+\epsilon_i,\]
where $\mu_j$, the median of $X_{ij}$, is introduced merely to
identify $\beta_0$ with the typical color
of the sample, and $\epsilon_i$ is the deviation 
from the relation due to the intrinsic scatter. We assume
that $\epsilon_i\sim {\cal N}(0,\sigma^2)$ independently of the other
parameters. Let us write $\beta_{p+1}=\sigma$ to remind that 
the scatter is also to be determined.

The observed variables, $x_{ij}$ and $y_i$, are related to the
true variables by 
\[x_{ij}=X_{ij}+\epsilon_{ij},\]
\[y_i=Y_i+\epsilon_{Yi},\]
where $\epsilon_{ij}\sim {\cal N}(0,\sigma_{ij}^2)$ and $\epsilon_{Yi}\sim {\cal N}(0,\sigma_{Yi}^2)$.
We obtain 
\[y_i-\beta_0-\sum_{j=1}^p\beta_j(x_{ij}-\mu_j)=\epsilon_{Yi}-\sum_{j=1}^p
\beta_j\epsilon_{ij}+\epsilon_i.\]
Assuming that $\epsilon_i$ is not correlated with the $\epsilon_{ij}$ 
and $\epsilon_{Yi}$, and that the covariances between $\epsilon_{ij}$
and $\epsilon_{ik}$  and between $\epsilon_{ij}$
and $\epsilon_{ijY}$ are $\gamma_{ijk}$ and $\gamma_{ijY}$, respectively,
we obtain that $y_i-\beta_0-\sum_{j=1}^p\beta_j(x_{ij}-\mu_j)\sim {\cal N}(0,{\sigma_i}^2+\beta_{p+1}^2)$
with 
\[{\sigma_i}^2=\sigma_{Yi}^2+\sum_{j=1}^p\beta_j^2\sigma_{ij}^2
-2\sum_{j=1}^p\beta_j\gamma_{ijY}-\sum_{j,k=1}^p\beta_j\beta_k\gamma_{ijk}.\]

We have tested two procedures to estimate the $\beta_j$ from our
sample. 
The first one estimates the $\beta_j$ by maximizing the logarithm of
the likelihood $\Lambda$ with respect to the $\beta_j, j\in[0,p+1]$, where
\begin{eqnarray*}
\ln(\Lambda)=-\frac{1}{2}\sum_{i=1}^n\Biggl\lbrace\!\!&\!\!&\!\!
\displaystyle{\frac{[y_i-\beta_0-\sum_{j=1}^p\beta_j
(x_{ij}-\mu_j)]^2}{{\sigma_i}^2
+\beta_{p+1}^2}}\\
\!\!&\!\!&\!\!\mbox{}+\ln({\sigma_i}^2+\beta_{p+1}^2)\Biggr\rbrace
\end{eqnarray*}
according to the formulae given above. The covariance matrix of the
$\beta_j$ is computed by inverting the curvature matrix 
$\left(\alpha_{ij}\right)$, $(i,j)\in[0,p+1]^2$, where
\[\alpha_{ij}=\frac{\partial^2\left[-\ln(\Lambda)\right]}{\partial\beta_i
\partial\beta_j}
\textrm{\qquad{}(Kendall \& Stuart \cite{KS}).}\]
Maximum likelihood (ML) estimators are often biased and we therefore
wish to compare to another method. 

The second procedure (MCES estimators) has been developed 
by Akritas \& Bershady (\cite{AB}) 
and Akritas (\cite{Akritas}). According to the authors, it yields unbiased
estimators.
However, the $\sigma_{Yi}$ are not used 
in the expression of the $\beta_i, i\le p$, which means that they have
all the same weight. This is especially not satisfying if the
intrinsic scatter is small or comparable to the uncertainties
in the colors -- which 
happens in particular for elliptical galaxies -- and may give an
excessive importance to outliers.
The authors also do not provide the intrinsic scatter
$\beta_{p+1}$ and  we estimate it from
\[
\beta_{p+1}^2=
\frac{1}{n}\sum_{i=1}^n\Bigl\lbrace\bigl[y_i-\beta_0
-\sum_{j=1}^p\beta_j\left(x_{ij}-\mu_j\right)\bigr]^2
-{\sigma_i}^2\Bigr\rbrace
\]
but when the intrinsic scatter is smaller or of the same order than 
the uncertainties in the observables, it is 
underestimated\footnote{This is shown by the fact that the ML and 
MCES estimators of the scatter in the effective colors 
(which are less uncertain) are similar and are also close
to the ML estimator for the total colors, whereas the MCES intrinsic
scatter in the total colors is smaller.} and we may
even obtain a negative (and
meaningless) value. We then assume $\beta_{p+1}=0$.
Uncertainties on the $\beta_j$ have been computed by a bootstrap 
with replacement on the $(x_{ij},y_i)$ rather than by using the
cumbersome formulae proposed by Akritas \& Bershady (\cite{AB})
and Akritas (\cite{Akritas}), which anyway are not available for $\beta_{p+1}$.

Note that the uncertainties determined either from the curvature matrix
or by bootstrap become themselves very uncertain when the number
of galaxies is less than about 20 to 30.

In the following, values determined by the ML and MCES estimators are
written in boldface and italic characters, respectively.

Because we are mainly interested in ``normal'' galaxies, we have to
reject a few galaxies with unusual colors (Buta et al. \cite{Buta}). 
Most of them are extremely
blue and, according to their optical colors (when available), 
are presumably starbursting galaxies. Some blue compact dwarfs
may also be misclassified from their morphology as ellipticals.
In the opposite,
our sample contains some galaxies with very red colors, especially in
the $K$-band and their NIR emission is dominated by an active nucleus.
To reject these galaxies, we apply an iterative method. 
Rather than using the standard deviation around the best fit, which is
very sensitive to the outliers, we estimate the observational scatter
from the median of the absolute deviations.
At each step, we compute 
\[\delta=1.48\mathop{\rm median}\limits_{i\in[1,n]}\biggl|y_i-\beta_0
-\sum_{j=1}^p\beta_j(x_{ij}-\mu_j)\biggr|\]
from the
sample (including the ``abnormal'' galaxies), the factor 1.48 ensuring that in the case of a perfectly 
Gaussian distribution, the standard deviation is $\delta$; 
we define ``normal'' galaxies as those having $\left|y_i-\beta_0
-\sum_{j=1}^p\beta_j(x_{ij}-\mu_j)\right|<3\delta$
and then determine the $\beta_j$ from them only 
by either of the methods discussed above.
We iterate this procedure 4 times, but the convergence is usually
achieved at the third one. 
\subsection{The type-color relation}
\begin{figure*}
\resizebox{!}{22cm}{\rotatebox{0}{\includegraphics{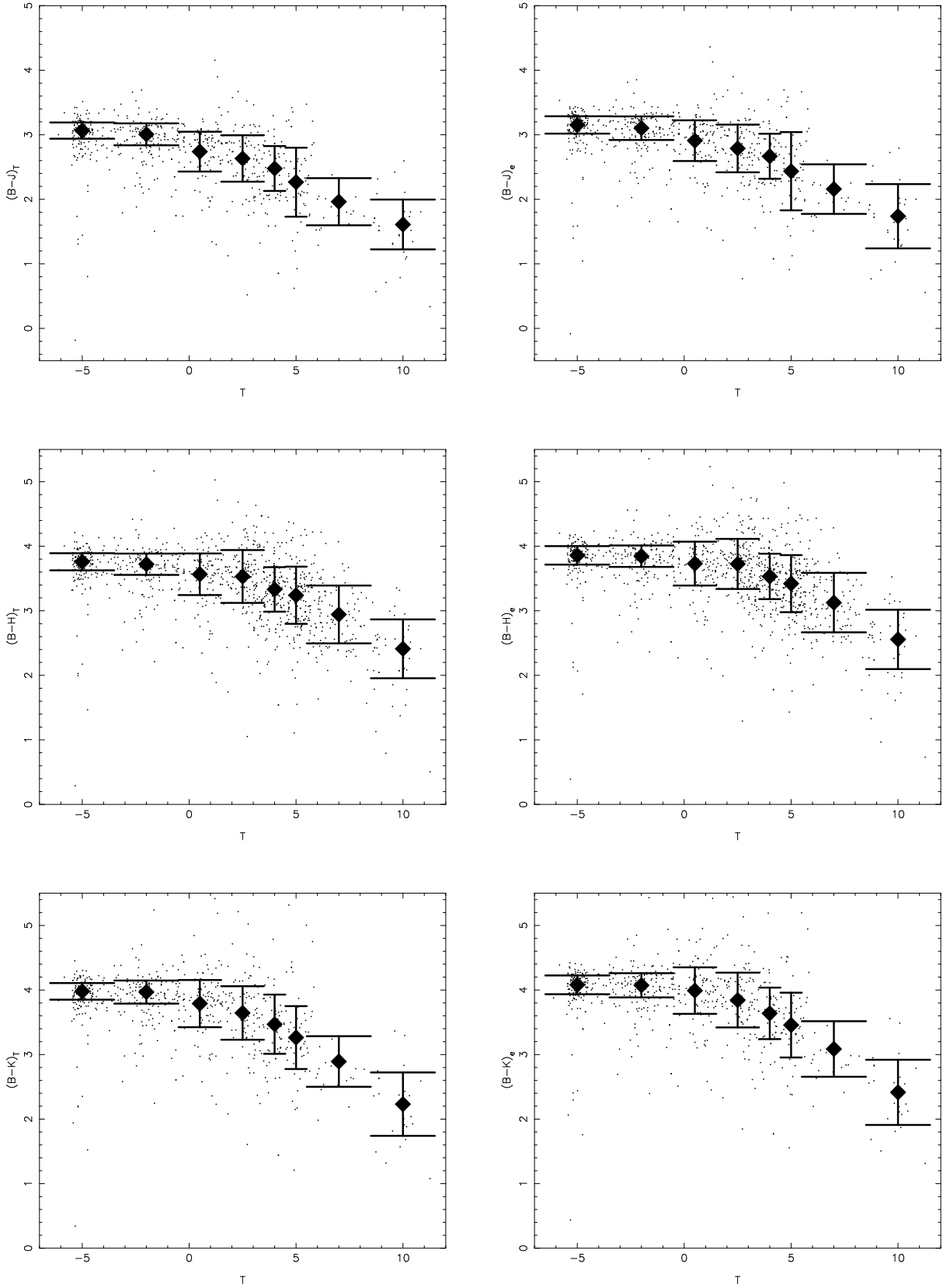}}}
\caption{Total and effective optical-to-NIR colors as a function
of the type. The diamonds are the standard colors determined by ML.
The vertical segments show the intrinsic scatter and the
horizontal ones the width of the type-bin.}
\label{figTBJHK}
\end{figure*}
The total and effective $B-J$, $B-H$ and $B-K$ colors
have been plotted as a function of the morphological type
on Figure~\ref{figTBJHK}. A very impressive trend to bluer colors 
with advancing type is observed. The mean $(B-H)_{\mathrm{T}}$ 
of irregular galaxies is 1.34 mag bluer
than the one of ellipticals, to be compared to 0.68 in $U-B$, 0.47 in
$B-V$, 0.46 in $V-I_{\mathrm{c}}$ and only 0.23 in $V-R_{\mathrm{c}}$
(Fioc \& Rocca-Volmerange, in preparation). A similar gap between the 
$(B-H)_{\mathrm{T}}$ of ellipticals and irregulars was 
obtained in $(B-H)_{\mathrm{e}}$ by Buta (\cite{Buta95}) 
from a sample of 225 galaxies.

The $(B-H)_{\mathrm{T}}$ colors determined by the ML estimator are
almost always systematically redder by $\sim 0.02$ mag
than those computed using the MCES estimator. This bias comes from the
fact that the uncertainties are smaller for brighter galaxies in the
NIR, which are also usually redder. The ML estimator
giving more weight to the data with smaller uncertainties, it is
biased to the red.

A significant scatter is also observed within each type.
Part of it is due to the observational uncertainties, but it comes
mainly from the {\em intrinsic} scatter in the colors.
The intrinsic scatter increases from $\mathbf{0.13}/\mathit{0.08}$, depending on the
estimator, for ellipticals to $\sim 0.4$ for Sb and remains nearly constant
at later types.
\begin{table*}
\begin{center}
\begin{tabular}{l|r||c|c||c|c}
type & $N$ & $\beta_0[(B-J)_{\mathrm{T}}]$& $\sigma$ &$\beta_0[(B-J)_{\mathrm{e}}]$ &$\sigma$
\\
\hline
E&140&$\mathbf{  3.07\pm  0.02}$&$\mathbf{  0.13\pm  0.02}$&$\mathbf{  3.15\pm  0.01}$&$\mathbf{  0.13\pm  0.01}$\\
&&$\mathit{  3.02\pm  0.02}$&$\mathit{  0.10\pm  0.03}$&$\mathit{  3.13\pm  0.02}$&$\mathit{  0.14\pm  0.02}$\\
S0&155&$\mathbf{  3.01\pm  0.02}$&$\mathbf{  0.17\pm  0.02}$&$\mathbf{  3.10\pm  0.02}$&$\mathbf{  0.18\pm  0.01}$\\
&&$\mathit{  2.97\pm  0.02}$&$\mathit{  0.15\pm  0.03}$&$\mathit{  3.07\pm  0.02}$&$\mathit{  0.24\pm  0.02}$\\
Sa& 97&$\mathbf{  2.74\pm  0.04}$&$\mathbf{  0.31\pm  0.03}$&$\mathbf{  2.91\pm  0.04}$&$\mathbf{  0.31\pm  0.03}$\\
&&$\mathit{  2.74\pm  0.04}$&$\mathit{  0.25\pm  0.05}$&$\mathit{  2.93\pm  0.03}$&$\mathit{  0.27\pm  0.03}$\\
Sb& 95&$\mathbf{  2.63\pm  0.04}$&$\mathbf{  0.36\pm  0.04}$&$\mathbf{  2.79\pm  0.04}$&$\mathbf{  0.37\pm  0.03}$\\
&&$\mathit{  2.62\pm  0.05}$&$\mathit{  0.33\pm  0.04}$&$\mathit{  2.79\pm  0.04}$&$\mathit{  0.36\pm  0.04}$\\
Sbc& 47&$\mathbf{  2.48\pm  0.06}$&$\mathbf{  0.35\pm  0.05}$&$\mathbf{  2.67\pm  0.06}$&$\mathbf{  0.35\pm  0.04}$\\
&&$\mathit{  2.45\pm  0.06}$&$\mathit{  0.17\pm  0.10}$&$\mathit{  2.66\pm  0.05}$&$\mathit{  0.26\pm  0.07}$\\
Sc& 46&$\mathbf{  2.27\pm  0.09}$&$\mathbf{  0.53\pm  0.07}$&$\mathbf{  2.44\pm  0.10}$&$\mathbf{  0.61\pm  0.07}$\\
&&$\mathit{  2.24\pm  0.08}$&$\mathit{  0.45\pm  0.10}$&$\mathit{  2.43\pm  0.10}$&$\mathit{  0.56\pm  0.08}$\\
Sd& 26&$\mathbf{  1.96\pm  0.09}$&$\mathbf{  0.37\pm  0.07}$&$\mathbf{  2.16\pm  0.09}$&$\mathbf{  0.38\pm  0.07}$\\
&&$\mathit{  1.93\pm  0.08}$&$\mathit{  0.23\pm  0.12}$&$\mathit{  2.14\pm  0.08}$&$\mathit{  0.32\pm  0.10}$\\
Im& 28&$\mathbf{  1.61\pm  0.09}$&$\mathbf{  0.39\pm  0.07}$&$\mathbf{  1.74\pm  0.10}$&$\mathbf{  0.50\pm  0.08}$\\
&&$\mathit{  1.52\pm  0.10}$&$\mathit{  0.41\pm  0.10}$&$\mathit{  1.72\pm  0.10}$&$\mathit{  0.47\pm  0.08}$\\
\end{tabular}
\caption{$(B-J)_{\mathrm{T}}$ and $(B-J)_{\mathrm{e}}$ colors per
type. The values computed with the ML and MCES estimators are written
respectively in boldface and italic characters. $\sigma$ is the
intrinsic scatter.}
\label{BmoinsJ}
\end{center}
\end{table*}
\begin{table*}
\begin{center}
\begin{tabular}{l|r||c|c||c|c}
type & $N$ & $\beta_0[(B-H)_{\mathrm{T}}]$& $\sigma$ &$\beta_0[(B-H)_{\mathrm{e}}]$ &$\sigma$
\\
\hline
E&142&$\mathbf{  3.76\pm  0.02}$&$\mathbf{  0.13\pm  0.01}$&$\mathbf{  3.86\pm  0.02}$&$\mathbf{  0.14\pm  0.01}$\\
&&$\mathit{  3.73\pm  0.02}$&$\mathit{  0.08\pm  0.04}$&$\mathit{  3.84\pm  0.02}$&$\mathit{  0.13\pm  0.02}$\\
S0&157&$\mathbf{  3.72\pm  0.02}$&$\mathbf{  0.17\pm  0.02}$&$\mathbf{  3.85\pm  0.02}$&$\mathbf{  0.17\pm  0.01}$\\
&&$\mathit{  3.69\pm  0.02}$&$\mathit{  0.14\pm  0.03}$&$\mathit{  3.82\pm  0.02}$&$\mathit{  0.19\pm  0.02}$\\
Sa&110&$\mathbf{  3.57\pm  0.04}$&$\mathbf{  0.32\pm  0.03}$&$\mathbf{  3.73\pm  0.04}$&$\mathbf{  0.34\pm  0.03}$\\
&&$\mathit{  3.54\pm  0.04}$&$\mathit{  0.31\pm  0.04}$&$\mathit{  3.73\pm  0.04}$&$\mathit{  0.33\pm  0.03}$\\
Sb&171&$\mathbf{  3.53\pm  0.03}$&$\mathbf{  0.41\pm  0.03}$&$\mathbf{  3.73\pm  0.03}$&$\mathbf{  0.39\pm  0.02}$\\
&&$\mathit{  3.52\pm  0.03}$&$\mathit{  0.40\pm  0.03}$&$\mathit{  3.71\pm  0.03}$&$\mathit{  0.40\pm  0.03}$\\
Sbc&110&$\mathbf{  3.33\pm  0.04}$&$\mathbf{  0.34\pm  0.03}$&$\mathbf{  3.53\pm  0.04}$&$\mathbf{  0.35\pm  0.03}$\\
&&$\mathit{  3.30\pm  0.03}$&$\mathit{  0.29\pm  0.04}$&$\mathit{  3.51\pm  0.03}$&$\mathit{  0.32\pm  0.03}$\\
Sc&125&$\mathbf{  3.24\pm  0.04}$&$\mathbf{  0.44\pm  0.03}$&$\mathbf{  3.42\pm  0.04}$&$\mathbf{  0.44\pm  0.03}$\\
&&$\mathit{  3.20\pm  0.04}$&$\mathit{  0.42\pm  0.04}$&$\mathit{  3.39\pm  0.04}$&$\mathit{  0.43\pm  0.03}$\\
Sd&135&$\mathbf{  2.94\pm  0.04}$&$\mathbf{  0.45\pm  0.03}$&$\mathbf{  3.13\pm  0.04}$&$\mathbf{  0.46\pm  0.03}$\\
&&$\mathit{  2.93\pm  0.04}$&$\mathit{  0.45\pm  0.03}$&$\mathit{  3.12\pm  0.04}$&$\mathit{  0.47\pm  0.03}$\\
Im& 42&$\mathbf{  2.41\pm  0.08}$&$\mathbf{  0.46\pm  0.06}$&$\mathbf{  2.56\pm  0.08}$&$\mathbf{  0.46\pm  0.06}$\\
&&$\mathit{  2.38\pm  0.08}$&$\mathit{  0.42\pm  0.07}$&$\mathit{  2.54\pm  0.08}$&$\mathit{  0.44\pm  0.06}$\\
\end{tabular}
\caption{$(B-H)_{\mathrm{T}}$ and $(B-H)_{\mathrm{e}}$ colors per type.}
\label{BmoinsH}
\end{center}
\end{table*}
\begin{table*}
\begin{center}
\begin{tabular}{l|r||c|c||c|c}
type & $N$ & $\beta_0[(B-K)_{\mathrm{T}}]$& $\sigma$ &$\beta_0[(B-K)_{\mathrm{e}}]$ &$\sigma$
\\
\hline
E&142&$\mathbf{  3.98\pm  0.02}$&$\mathbf{  0.13\pm  0.01}$&$\mathbf{  4.08\pm  0.02}$&$\mathbf{  0.15\pm  0.01}$\\
&&$\mathit{  3.95\pm  0.02}$&$\mathit{  0.09\pm  0.04}$&$\mathit{  4.06\pm  0.02}$&$\mathit{  0.14\pm  0.02}$\\
S0&159&$\mathbf{  3.97\pm  0.02}$&$\mathbf{  0.18\pm  0.02}$&$\mathbf{  4.07\pm  0.02}$&$\mathbf{  0.19\pm  0.01}$\\
&&$\mathit{  3.95\pm  0.02}$&$\mathit{  0.15\pm  0.03}$&$\mathit{  4.06\pm  0.02}$&$\mathit{  0.19\pm  0.02}$\\
Sa& 99&$\mathbf{  3.79\pm  0.04}$&$\mathbf{  0.37\pm  0.03}$&$\mathbf{  3.99\pm  0.04}$&$\mathbf{  0.36\pm  0.03}$\\
&&$\mathit{  3.79\pm  0.04}$&$\mathit{  0.33\pm  0.04}$&$\mathit{  4.00\pm  0.04}$&$\mathit{  0.35\pm  0.03}$\\
Sb&100&$\mathbf{  3.64\pm  0.05}$&$\mathbf{  0.41\pm  0.04}$&$\mathbf{  3.84\pm  0.05}$&$\mathbf{  0.42\pm  0.04}$\\
&&$\mathit{  3.62\pm  0.05}$&$\mathit{  0.42\pm  0.05}$&$\mathit{  3.82\pm  0.05}$&$\mathit{  0.43\pm  0.04}$\\
Sbc& 48&$\mathbf{  3.47\pm  0.08}$&$\mathbf{  0.46\pm  0.06}$&$\mathbf{  3.64\pm  0.07}$&$\mathbf{  0.40\pm  0.05}$\\
&&$\mathit{  3.45\pm  0.07}$&$\mathit{  0.32\pm  0.10}$&$\mathit{  3.64\pm  0.06}$&$\mathit{  0.32\pm  0.06}$\\
Sc& 46&$\mathbf{  3.26\pm  0.09}$&$\mathbf{  0.49\pm  0.07}$&$\mathbf{  3.46\pm  0.09}$&$\mathbf{  0.50\pm  0.06}$\\
&&$\mathit{  3.22\pm  0.08}$&$\mathit{  0.39\pm  0.10}$&$\mathit{  3.43\pm  0.08}$&$\mathit{  0.45\pm  0.07}$\\
Sd& 24&$\mathbf{  2.89\pm  0.10}$&$\mathbf{  0.39\pm  0.08}$&$\mathbf{  3.09\pm  0.10}$&$\mathbf{  0.43\pm  0.08}$\\
&&$\mathit{  2.80\pm  0.08}$&$\mathit{  0.21\pm  0.13}$&$\mathit{  3.01\pm  0.09}$&$\mathit{  0.31\pm  0.10}$\\
Im& 25&$\mathbf{  2.23\pm  0.11}$&$\mathbf{  0.49\pm  0.09}$&$\mathbf{  2.41\pm  0.11}$&$\mathbf{  0.50\pm  0.08}$\\
&&$\mathit{  2.21\pm  0.11}$&$\mathit{  0.45\pm  0.10}$&$\mathit{  2.40\pm  0.11}$&$\mathit{  0.50\pm  0.08}$\\
\end{tabular}
\caption{$(B-K)_{\mathrm{T}}$ and $(B-K)_{\mathrm{e}}$ colors per type.}
\label{BmoinsK}
\end{center}
\end{table*}
\subsection{The $\Rvc$-color relation}
\label{corrinclin}
The color-inclination relation is potentially a very powerful constraint on
the amount of dust and its distribution relatively to stars.
As a disk becomes more and more inclined, its optical depth increases
and the colors redden. This must be especially striking when
one of the band (e.g. $B$) is heavily extinguished whereas the dust
is almost transparent in the other one (e.g. $H$). For these reasons,
various studies have tried to determine a color-inclination relation, 
regrouping the galaxies either as a function of type (e.g. Boselli \&
Gavazzi \cite{BG94}) or as a function 
of their NIR absolute magnitude (Tully et al. \cite{Tully}). 
The samples used in these studies
where however small (about 100 galaxies) and it is worth to 
look at the relation once again with our much larger catalog.

For an oblate ellipsoid, which is a standard model for a galaxy
disk (Hubble \cite{Hubble}), 
the inclination $i$ (face-on corresponds to $i=0$)
is related to the true ratio $q_0$ of the minor to the major axis
and to the apparent ratio $q\simeq 1/\Rvc$ by
\[\cos^2i=\frac{q^2-q_0^2}{1-q_0^2}.\]

Practically, we prefer to establish a relation between the colors
and $\logd(\Rvc)$, both because $\Rvc$ is the directly observed 
quantity and because the linear estimators used here provide
a better fit when $\logd(\Rvc)$ is used rather than $i$ or $\cos(i)$.
\begin{figure*}
\resizebox{!}{22cm}{\rotatebox{0}{\includegraphics{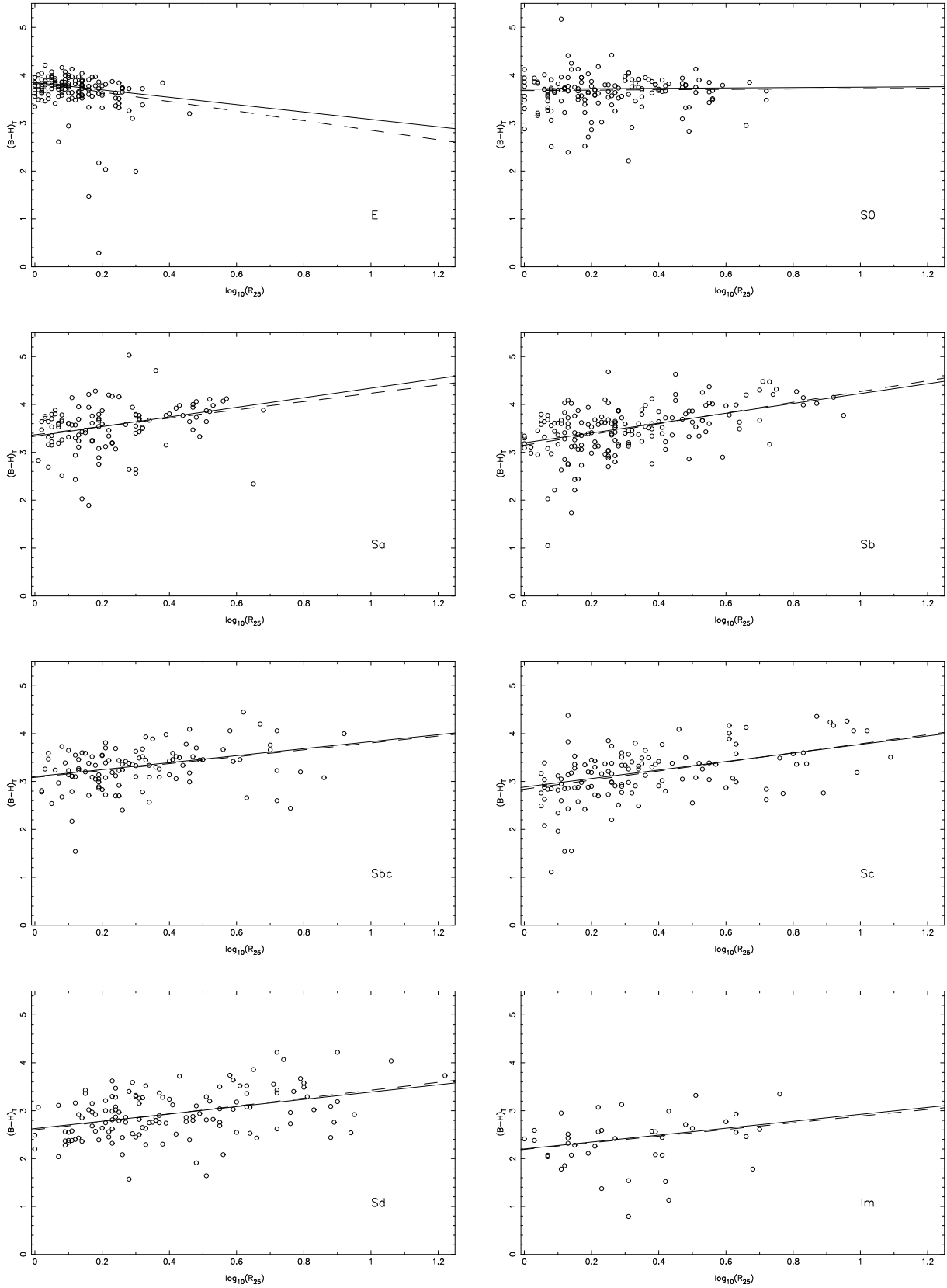}}}
\caption{$(B-H)_{\mathrm{T}}$-$\logd(\Rvc)$ relation per type.
The solid and dashed lines correspond respectively to the ML and
MCES estimators.} 
\label{figinclin}
\end{figure*}

\begin{figure}
\resizebox{\hsize}{!}{\rotatebox{0}{\includegraphics{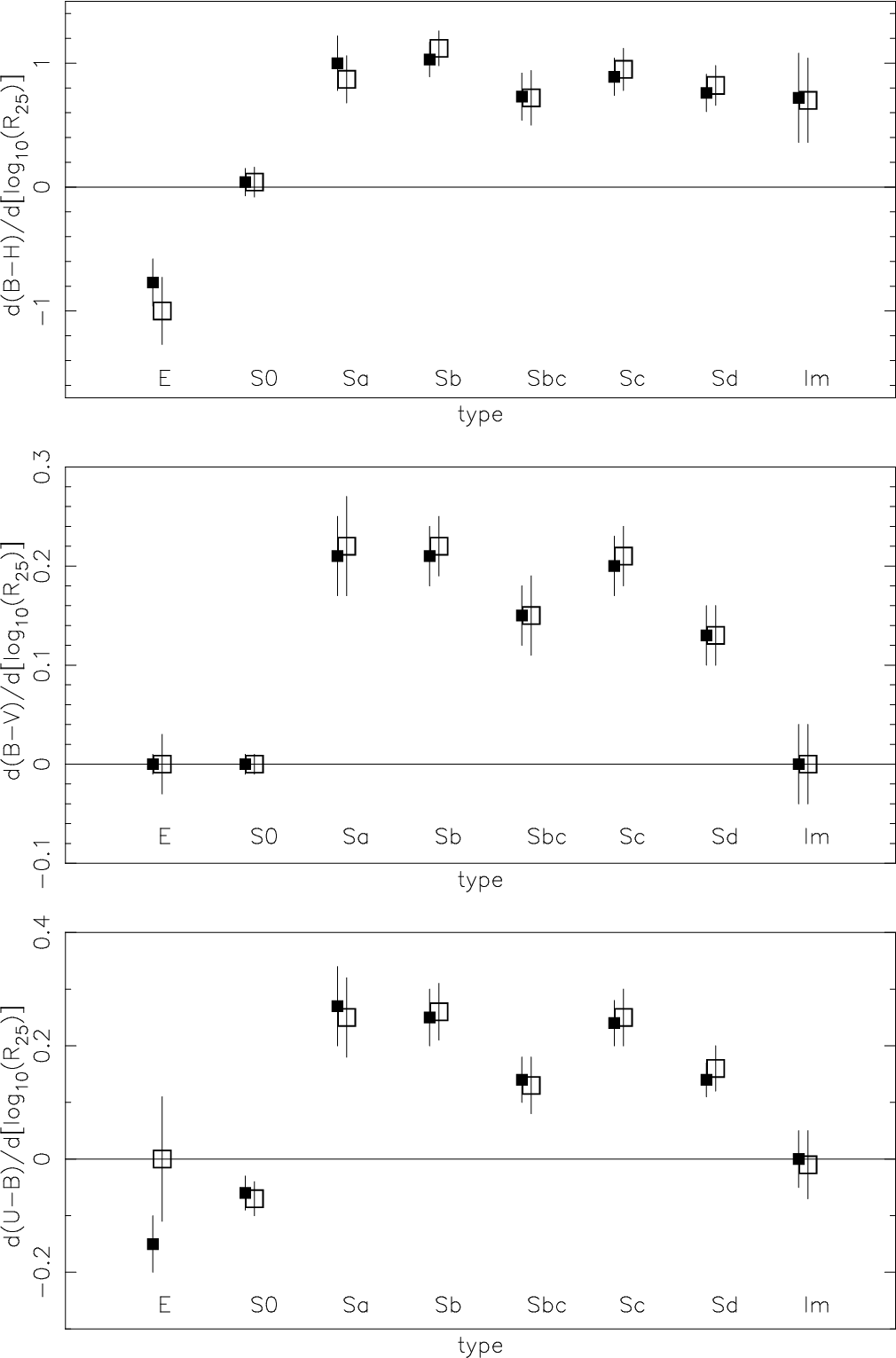}}}.
\caption{Slopes of the $(B-H)_{\mathrm{T}}$-, $(B-V)_{\mathrm{T}}$-,
and $(U-B)_{\mathrm{T}}$-$\logd(\Rvc)$ relations as a function of type. 
The black and white squares correspond respectively 
to the ML and MCES estimators.}
\label{couleurs_R25}
\end{figure}

The relations are plotted in Fig.~\ref{figinclin} and 
statistical estimators of the slope and the color are given 
in table~\ref{tabinclin} for $B-H$.
\begin{table*}
\begin{center}
\begin{tabular}{l|r|c|c|c|c}
type & $N$ &$\beta_0[(B-H)_{\mathrm{T}}]$&$\beta_1[\logd(\Rvc)]$&$\mu_1$&$\sigma$\\
\hline
E&138&$\mathbf{  3.77\pm  0.02}$&$\mathbf{ -0.77\pm  0.19}$&0.10&$\mathbf{  0.12\pm  0.01}$\\
&&$\mathit{  3.75\pm  0.02}$&$\mathit{ -1.00\pm  0.27}$&&$\mathit{  (0.03\pm  0.03)}$\\
S0&153&$\mathbf{  3.72\pm  0.02}$&$\mathbf{  0.04\pm  0.11}$&0.20&$\mathbf{  0.17\pm  0.02}$\\
&&$\mathit{  3.69\pm  0.02}$&$\mathit{  0.04\pm  0.12}$&&$\mathit{  0.15\pm  0.03}$\\
Sa&110&$\mathbf{  3.53\pm  0.04}$&$\mathbf{  1.00\pm  0.22}$&0.19&$\mathbf{  0.31\pm  0.03}$\\
&&$\mathit{  3.53\pm  0.04}$&$\mathit{  0.87\pm  0.19}$&&$\mathit{  0.24\pm  0.04}$\\
Sb&171&$\mathbf{  3.48\pm  0.03}$&$\mathbf{  1.03\pm  0.14}$&0.27&$\mathbf{  0.34\pm  0.02}$\\
&&$\mathit{  3.45\pm  0.03}$&$\mathit{  1.12\pm  0.14}$&&$\mathit{  0.34\pm  0.03}$\\
Sbc&110&$\mathbf{  3.29\pm  0.04}$&$\mathbf{  0.73\pm  0.19}$&0.25&$\mathbf{  0.34\pm  0.03}$\\
&&$\mathit{  3.27\pm  0.03}$&$\mathit{  0.72\pm  0.22}$&&$\mathit{  0.28\pm  0.04}$\\
Sc&125&$\mathbf{  3.14\pm  0.04}$&$\mathbf{  0.89\pm  0.15}$&0.29&$\mathbf{  0.37\pm  0.03}$\\
&&$\mathit{  3.12\pm  0.04}$&$\mathit{  0.95\pm  0.17}$&&$\mathit{  0.32\pm  0.03}$\\
Sd&135&$\mathbf{  2.89\pm  0.04}$&$\mathbf{  0.76\pm  0.15}$&0.34&$\mathbf{  0.40\pm  0.03}$\\
&&$\mathit{  2.88\pm  0.04}$&$\mathit{  0.82\pm  0.16}$&&$\mathit{  0.38\pm  0.03}$\\
Im& 41&$\mathbf{  2.41\pm  0.08}$&$\mathbf{  0.72\pm  0.36}$&0.29&$\mathbf{  0.40\pm  0.06}$\\
&&$\mathit{  2.39\pm  0.07}$&$\mathit{  0.70\pm  0.34}$&&$\mathit{  0.35\pm  0.08}$\\
\end{tabular}
\caption{$(B-H)_{\mathrm{T}}$ color per type as a function of $X_{i1}=\logd(\Rvc)$.}
\label{tabinclin}
\end{center}
\end{table*}
The slopes are represented as a function of type in
figure~\ref{couleurs_R25}.
Also shown on this graph are the slopes derived from the larger
$B-V$ and $U-B$ samples.

The typical slope of the $(B-H)_{\mathrm{T}}$-$\logd(\Rvc)$ relation is about 1 for spiral
galaxies, to be compared to 0.2--0.3 in $B-V$ and $U-B$.

Surprisingly high is the value of the slope for Sa galaxies. However,
the slope of the $(B-H)_{\mathrm{T}}$-{\em inclination} relation 
would be shallower
for Sa since $q_0$ is a decreasing function of type (Bottinelli et
al. \cite{Bottinelli}), 
i.e. Sa galaxies are thicker than later types, as also indicated by their
smaller median $\logd(\Rvc)$.

\begin{figure}
\resizebox{\hsize}{!}{\rotatebox{0}{\includegraphics{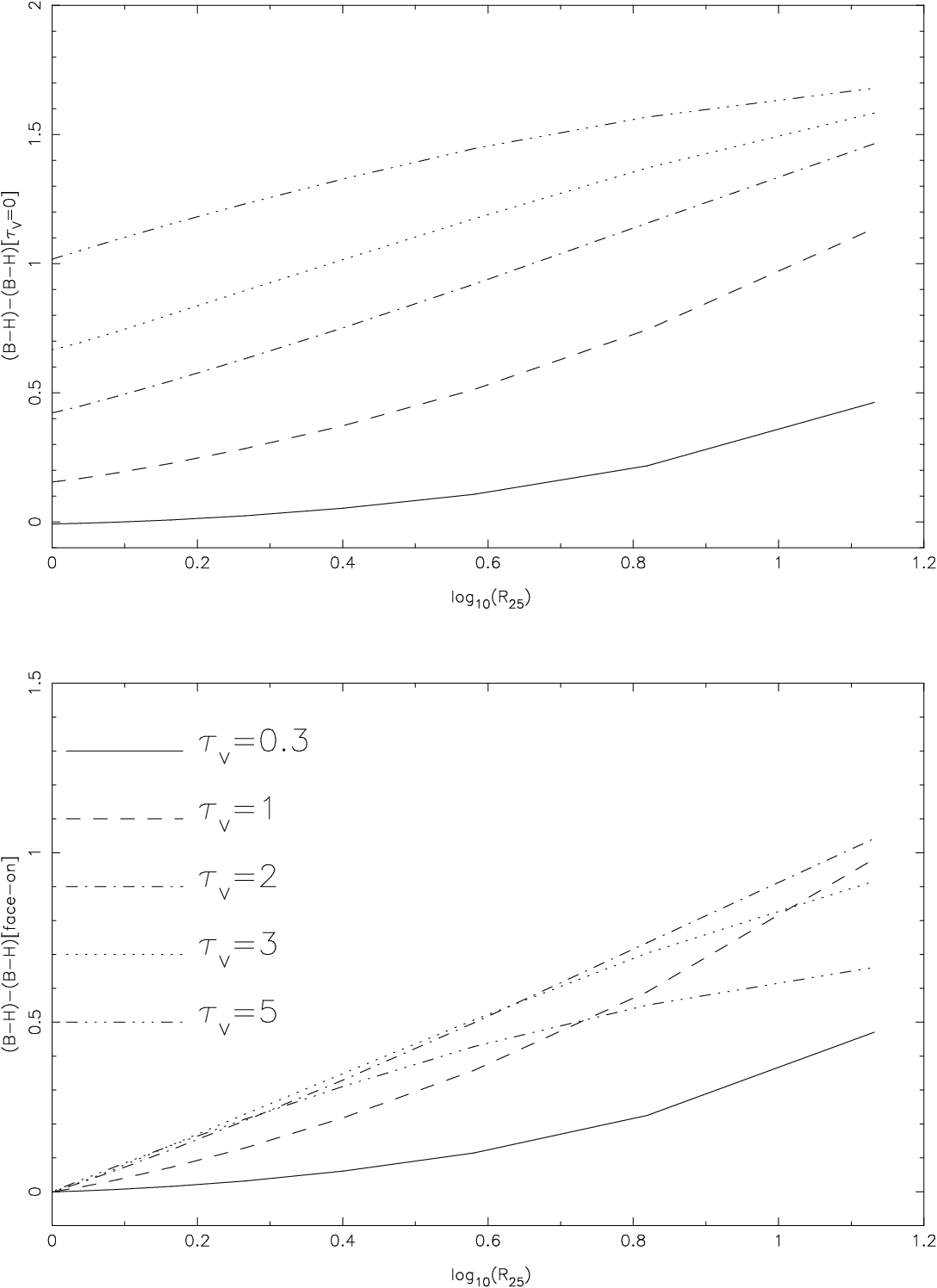}}}.
\caption{$(B-H)$-$\logd(\Rvc)$ relations computed with the radiative
transfer code (see text) for different face-on $V$-band optical
depths. We assume $q_0=0.18$, a typical value for an Sb.
{\bf Top:} reddening relative to the dust-free case. {\bf Bottom:} reddening
relative to face-on.}
\label{inclinBmoinsH}
\end{figure}

A dip is observed for Sbc galaxies, which seems strange since the
maximum extinction is expected precisely for this type (Fioc \&
Rocca-Volmerange \cite{FRV}). Though not inconsistent with a constant
slope for all spirals, given the uncertainties,
a similar dip is also obvious in the
slopes of the $(B-V)_{\mathrm{T}}$- and $(U-B)_{\mathrm{T}}$-$\logd(\Rvc)$ relations, suggesting
that this is a real phenomenon. One should first remember that 
the slope is not a measure of the total extinction but of the
{\em reddening relative to face-on}.
With a simple radiative transfer model where the stars and the dust
are distributed homogeneously in an infinite slab, we find that when 
the face-on optical depth $\tau_{\mathrm{V}}$ in the $V$-band
increases, 
the slope begins first to steepen (Fig.~\ref{inclinBmoinsH}b) 
but when $\tau_{\mathrm{V}}$ becomes larger than 1--2, the slope
flattens although the colors go on reddening (Fig.~\ref{inclinBmoinsH}a).
Although this modeling is simplistic and we can not infer from it the
value of the optical depth,
this suggests that the extinction may indeed be higher for Sbc
and that the slope has begun to decrease.
This conclusion is reinforced by the fact that the slope
{\em decreases} when we consider {\em effective} $B-H$ colors 
($\beta_1[\logd(\Rvc)]=\mathbf{0.60}/\mathit{0.55}$), which
are probably more extinguished than total colors.

We observe a positive slope for irregular galaxies in $B-H$, contrary
to $B-V$ and $U-B$ where none is detected. However, the
uncertainties are large and the null slope is within 2$\sigma$.
Moreover, our ``Im'' type regroups in fact not only Im galaxies ($T=10$)
but also Sm ($T=9$) which are more elongated and redder. 
The small size of the ``Im'' sample makes it impossible to 
decide whether this reddening is due to the dust or to intrinsically
redder populations in Sm galaxies.

No slope is observed for lenticular galaxies, which may indicate
that no significant amount of dust is present in S0 (see also
Sandage \& Visvanatan \cite{SV}). Another
possibility is that
the disk is overwhelmed by the bulge. No
inclination dependence of the colors would then be expected
if the bulge is spherically symmetric or if it is devoid of dust.

The most striking result is the negative slope observed for
ellipticals. This certainly does not come from the dust because
the slope would be positive. For ellipticals, $\Rvc$ is not directly
related to the inclination because their true shape is unknown.
In the mean, rounder ellipticals seem redder and are probably more
metal-rich (Terlevich et al. \cite{Terlevich}). They also tend
to be brighter (Tremblay \& Merritt \cite{TM}) which may link
the color-$\logd(\Rvc)$ relation of ellipticals to their
color-magnitude relation. 
No such relation is observed in $B-V$, 
which is anyway a poor indicator of the metallicity.
The $(U-B)_{\mathrm{T}}$-$\logd(\Rvc)$ is not very conclusive:
whereas the ML estimator provides a negative slope,
as in $B-H$, the MCES estimator does not detect any.
Since no obvious bias between the two estimators is observed 
for the other types and the intrinsic scatter is very small compared
to the uncertainties on the colors for
ellipticals -- which is the worst case for the MCES estimators --,
we tend to trust more the result given by the ML.  
\subsection{The color-magnitude relation}
\subsubsection{Global relation}
\begin{figure*}
\resizebox{!}{22cm}{\rotatebox{0}{\includegraphics{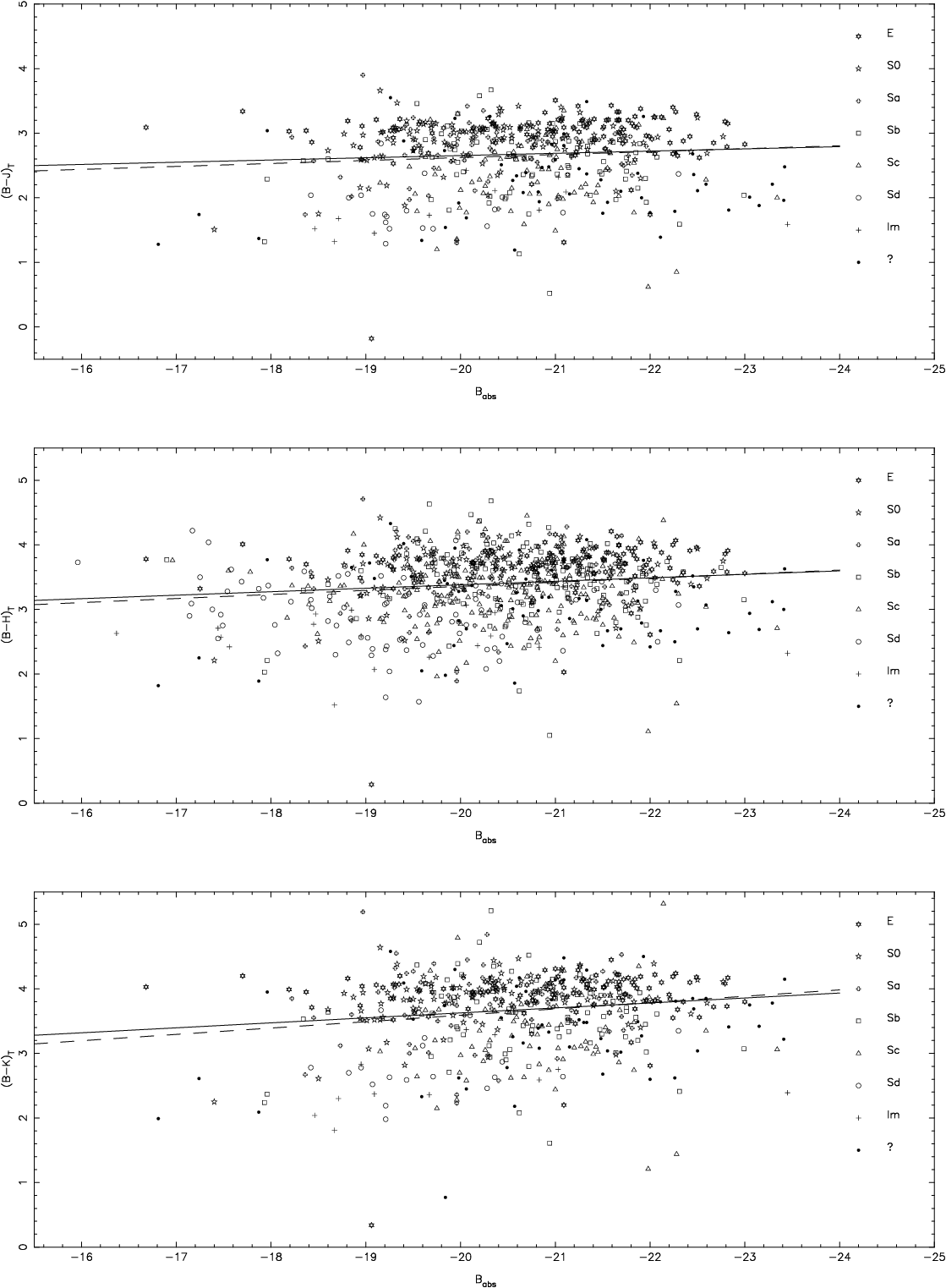}}}
\caption{Global $(B-\NIR)_{\mathrm{T}}$-absolute $B$ magnitude relations.}
\label{CMglobB}
\end{figure*}

\begin{figure*}
\resizebox{!}{22cm}{\rotatebox{0}{\includegraphics{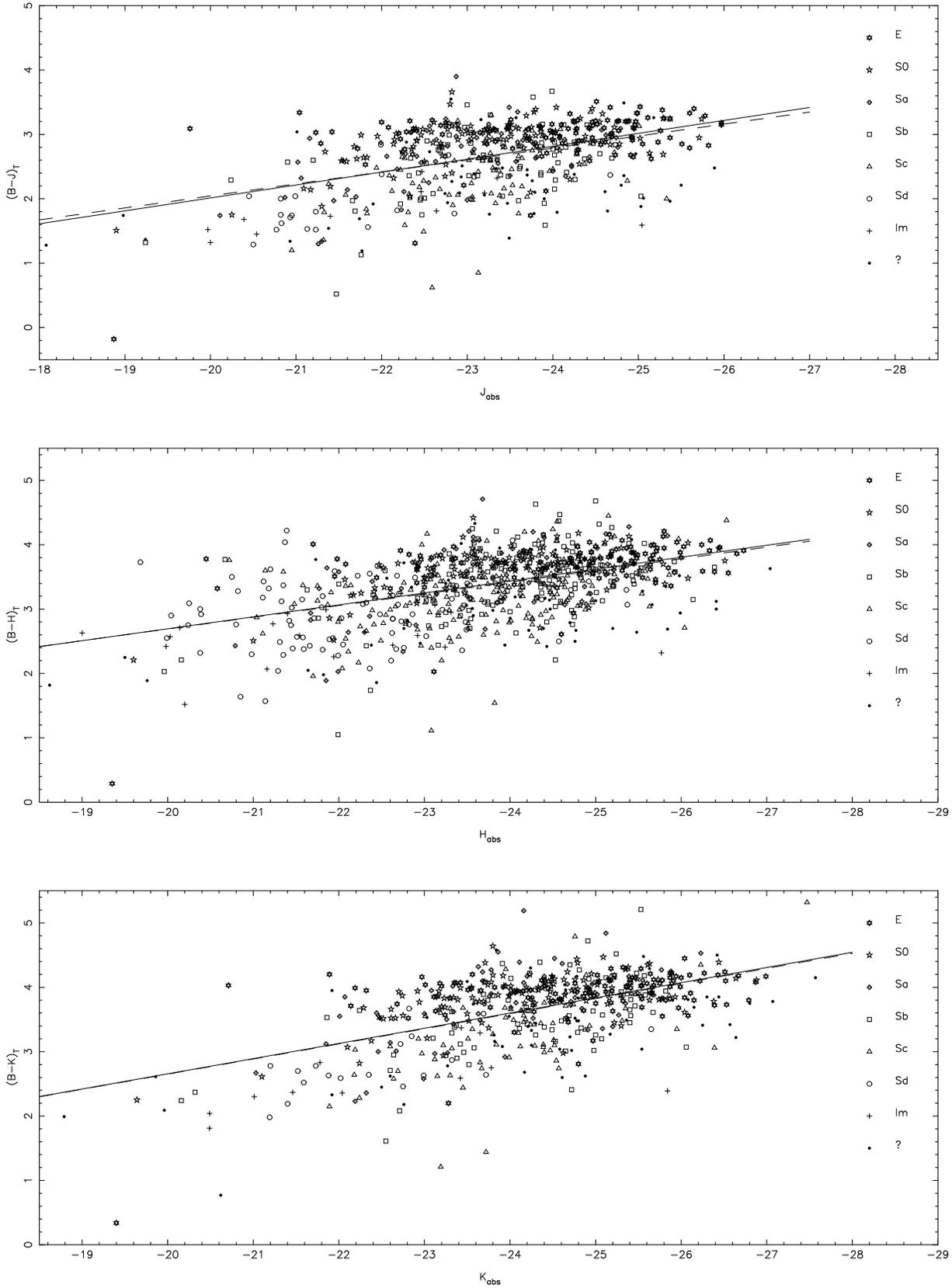}}}
\caption{Global $(B-\NIR)_{\mathrm{T}}$-absolute $\NIR$ magnitude relations.}
\label{CMglobNIR}
\end{figure*}

The color-magnitude (CM) relation is an important constraint on the
dynamical models of galaxy formation for the absolute magnitude
may be related to the mass of the galaxy via a mass-to-luminosity
ratio.
An important question is the choice of the reference band, optical
or NIR, used for the absolute magnitude. Usually, the NIR has been
adopted because it suffers little extinction. Another reason is that it is 
produced mainly by old giants and is therefore
expected to be a better indicator of the mass than the optical,
which is more sensitive to the recent star formation.
We have plotted on Figs.~\ref{CMglobB} and \ref{CMglobNIR} the color-absolute
magnitude relation with either the optical or the NIR as reference
band. The corresponding estimators are given in Table~\ref{tabCMglob}.
The $(B-\NIR)_{\mathrm{T}}$-$\NIR_{\mathrm{abs}}$ relation is much steeper than the
$(B-\NIR)_{\mathrm{T}}$-$B_{\mathrm{abs}}$ one.
If there was no intrinsic scatter, the slopes should be nearly the
same.
For example, if $(B-\NIR)_{\mathrm{T}}=\alpha\NIR_{\mathrm{abs}}+\kappa$, then
we should have
$(B-\NIR)_{\mathrm{T}}=[\alpha/(1+\alpha)] (B_{\mathrm{abs}}+\kappa)$, which has 
almost the same slope since $|\alpha|<<1$.
The scatter is moreover smaller when the NIR is the reference
band. To detect a color-magnitude relation, the choice of the NIR as
reference band is thus clearly favored.
\begin{table*}
\begin{center}
\begin{tabular}{@{\ }l@{\ }|@{\ }r@{\ }||@{\ }c@{\ }|@{\ }c@{\ }|@{\ }c@{\ }|@{\ }c@{\ }||@{\ }c@{\ }|@{\ }c@{\ }|@{\ }c@{\ }|@{\ }c@{\ }}
$(B-\NIR)_{\mathrm{T}}$ & $N$ & $\beta_0[(B-\NIR)_{\mathrm{T}}]$&$\beta_1(B_{\mathrm{abs}})$&$\mu_1$ &$\sigma$ &$\beta_0[(B-\NIR)_{\mathrm{T}}]$ &$\beta_1(\NIR_{\mathrm{abs}})$&$\mu_1$&$\sigma$\\
\hline
$(B-J)_{\mathrm{T}}$& 540&$\mathbf{  2.68\pm  0.02}$&$\mathbf{ -0.03\pm  0.02}$&-20.67&$\mathbf{  0.47\pm  0.02}$&$\mathbf{  2.68\pm  0.02}$&$\mathbf{ -0.20\pm  0.02}$&-23.34&$\mathbf{  0.41\pm  0.01}$\\
&&$\mathit{  2.65\pm  0.02}$&$\mathit{ -0.05\pm  0.03}$&&$\mathit{  0.46\pm  0.02}$&$\mathit{  2.67\pm  0.02}$&$\mathit{ -0.19\pm  0.02}$&&$\mathit{  0.40\pm  0.02}$\\
$(B-H)_{\mathrm{T}}$& 814&$\mathbf{  3.41\pm  0.02}$&$\mathbf{ -0.05\pm  0.01}$&-20.46&$\mathbf{  0.45\pm  0.01}$&$\mathbf{  3.42\pm  0.02}$&$\mathbf{ -0.19\pm  0.01}$&-23.90&$\mathbf{  0.40\pm  0.01}$\\
&&$\mathit{  3.39\pm  0.02}$&$\mathit{ -0.06\pm  0.02}$&&$\mathit{  0.45\pm  0.01}$&$\mathit{  3.40\pm  0.01}$&$\mathit{ -0.18\pm  0.01}$&&$\mathit{  0.40\pm  0.01}$\\
$(B-K)_{\mathrm{T}}$& 546&$\mathbf{  3.68\pm  0.02}$&$\mathbf{ -0.08\pm  0.02}$&-20.67&$\mathbf{  0.49\pm  0.02}$&$\mathbf{  3.69\pm  0.02}$&$\mathbf{ -0.24\pm  0.01}$&-24.38&$\mathbf{  0.41\pm  0.01}$\\
&&$\mathit{  3.66\pm  0.02}$&$\mathit{ -0.10\pm  0.03}$&&$\mathit{  0.49\pm  0.02}$&$\mathit{  3.68\pm  0.02}$&$\mathit{ -0.23\pm  0.02}$&&$\mathit{  0.40\pm  0.01}$\\
\end{tabular}
\caption{$(B-\NIR)_{\mathrm{T}}$ vs. $B_{\mathrm{abs}}$ or $\NIR_{\mathrm{abs}}$
color-magnitude relations, where $\NIR=J$, $H$ or $K$.}
\label{tabCMglob}
\end{center}
\end{table*}
Though the overall agreement is very good, we note that the ML
estimator of the slope is slightly biased when
compared to the {\em a priori} unbiased MCES estimator. Its slope is
``more positive'' in the $(B-\NIR)_{\mathrm{T}}$-$B_{\mathrm{abs}}$ relation because of the
positive correlation of the errors in $(B-\NIR)_{\mathrm{T}}$ and $B_{\mathrm{abs}}$,
and ``more negative'' in the $(B-\NIR)_{\mathrm{T}}$-$\NIR_{\mathrm{abs}}$ relation because of the
negative correlation of the errors in $(B-\NIR)_{\mathrm{T}}$ and $\NIR_{\mathrm{abs}}$.

The slope of the 
$(B-\NIR)_{\mathrm{T}}$-absolute magnitude relation
increases with the wavelength of the NIR band
for there is also a CM relation in the NIR colors. 
This effect
is less apparent in $(B-H)_{\mathrm{T}}$ because the $B-H$ sample contains
many late-type galaxies which have a flatter slope:
a linear fit for all the galaxies together 
is hence too simplistic.
\subsubsection{Relation per type}
The global color-magnitude relation discussed above is actually
dominated by star-forming galaxies
($\beta_1(H_{\mathrm{abs}})=\mathbf{-0.21}/\mathit{-0.20}$
against $\mathbf{-0.07}/\mathit{-0.05}$ for E/S0). 
These values are close to the slopes of the 
$(B_{\mathrm{j}}-K)[\Dvc]$-$K_{\mathrm{abs}}$ relations 
($-0.24\pm 0.03$ for Sa--Sdm and $-0.09\pm
0.04$ for E/S0)
determined by Mobasher et al. (\cite{Mobasher}).
The slope for spiral galaxies is much steeper than the
relations obtained by Bershady (\cite{Bershady}) for its 
spectral types bk, bm, am and fm.
This is however not surprising, because using {\em spectral}
types rather than {\em photometric} types bins galaxies as a function
of their colors and tends to suppress any color-magnitude relation.

\begin{figure}
\resizebox{\hsize}{!}{\rotatebox{-90}{\includegraphics{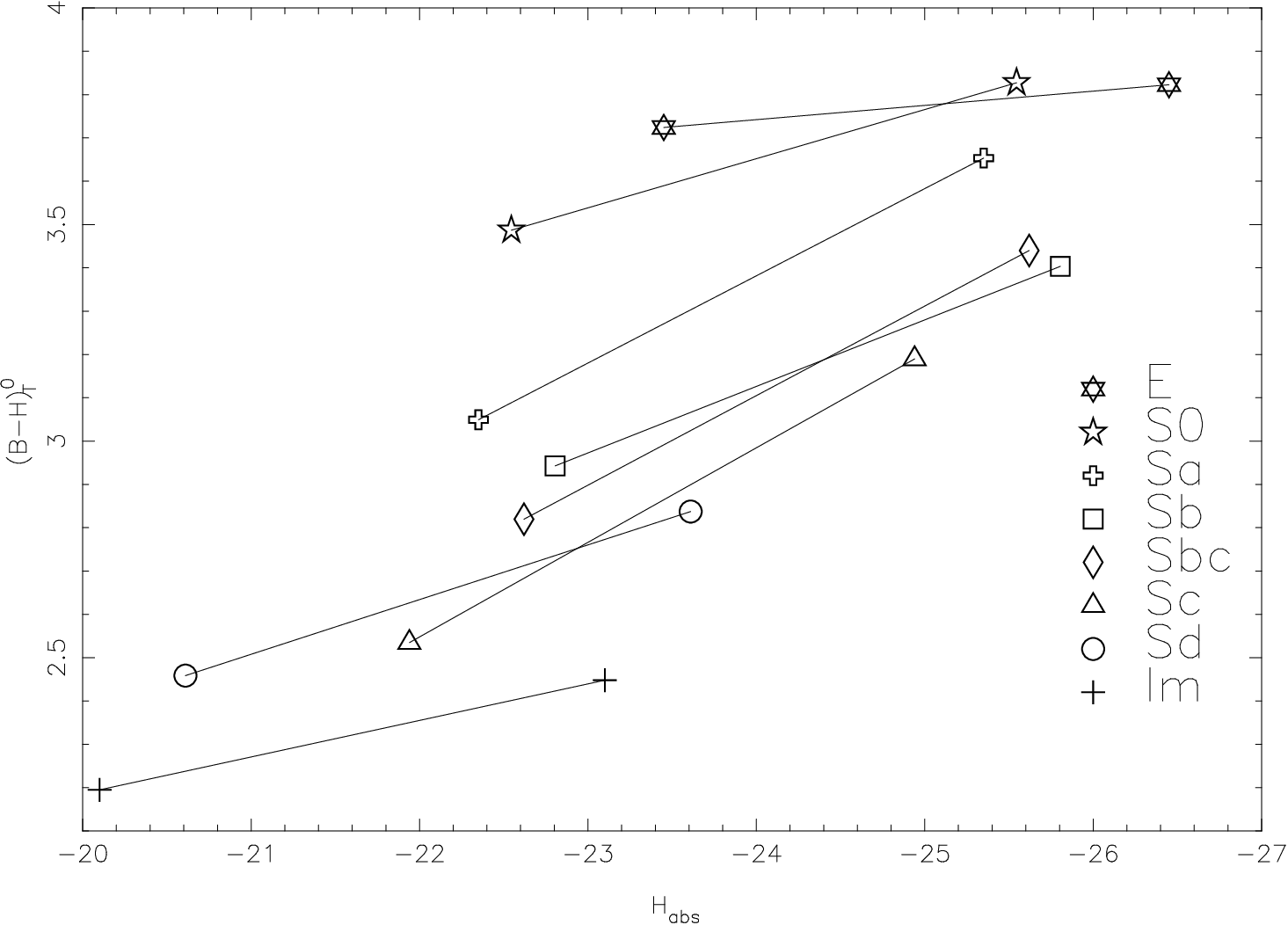}}}
\caption{Analytical $(B-H)^0_{\mathrm{T}}$-absolute $H$ magnitude 
relations. The segments correspond to the mean of the ML and MCES fits and
extend from $\mu_0(H_{\mathrm{abs}})+1.5$ 
to $\mu_0(H_{\mathrm{abs}})-1.5$ for each type.}
\label{segments}
\end{figure}

An especially interesting question is whether the mass is the main driving
force of galaxy evolution and star formation, or if the galaxy
type has also to be considered. To answer this question, it is worth to
look at the color-magnitude relation per type more thoroughly.
\begin{figure*}
\resizebox{!}{22cm}{\rotatebox{0}{\includegraphics{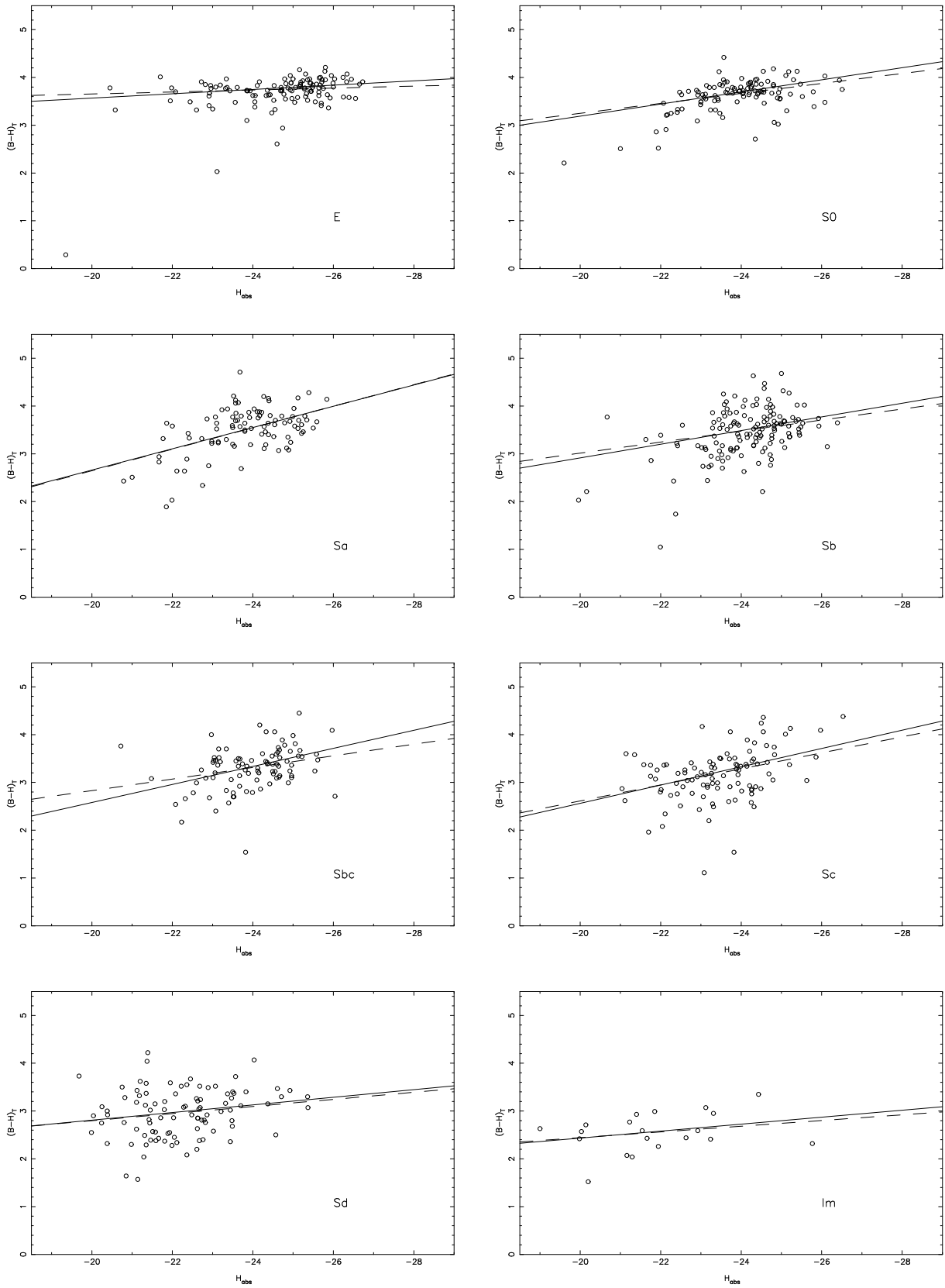}}}
\caption{$(B-H)_{\mathrm{T}}$-absolute $H$ magnitude
relations per type.}
\label{figCMtypeNIR}
\end{figure*}
\begin{figure*}
\resizebox{!}{22cm}{\rotatebox{0}{\includegraphics{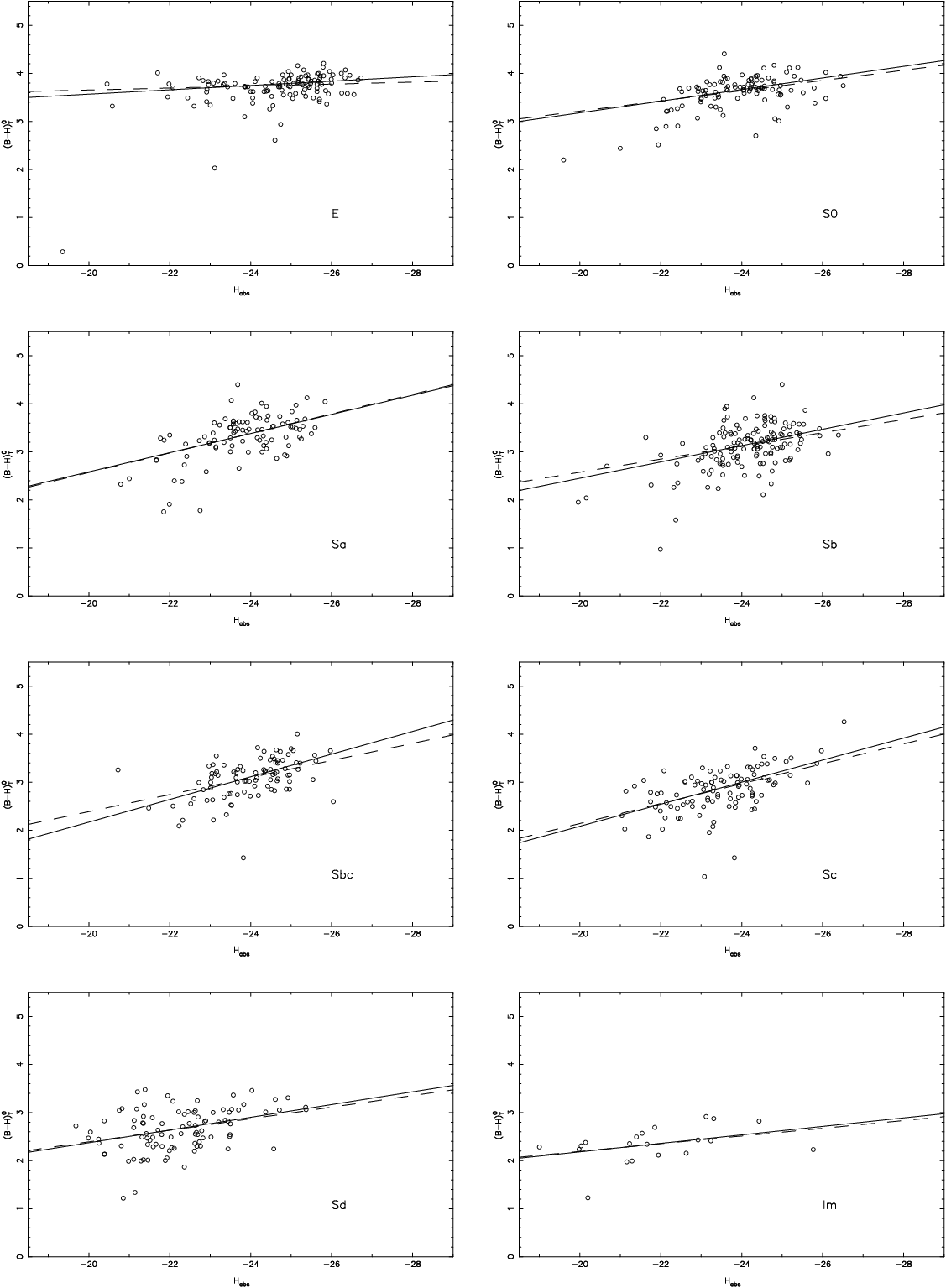}}}
\caption{$(B-H)^0_{\mathrm{T}}$-absolute $H$ magnitude
relations per type.}
\label{figCMtypeNIRcorr}
\end{figure*}
\begin{table*}
\begin{center}							 
\begin{tabular}{l|r|c|c|c|c||c|c|c}			
type & $N$ &$\beta_0[(B-H)_{\mathrm{T}}]$&$\beta_1(H_{\mathrm{abs}})$&$\mu_1$&$\sigma$&$\beta_0[(B-H)^0_{\mathrm{T}}]$&$\beta_1(H_{\mathrm{abs}})$&$\sigma$\\
\hline	 
E&119&$\mathbf{  3.79\pm  0.02}$&$\mathbf{ -0.05\pm  0.01}$&-24.95&$\mathbf{  0.12\pm  0.02}$&$\mathbf{  3.79\pm  0.02}$&$\mathbf{ -0.05\pm  0.01}$&$\mathbf{  0.12\pm  0.02}$\\
&&$\mathit{  3.76\pm  0.02}$&$\mathit{ -0.02\pm  0.02}$&&$\mathit{  (0.06\pm  0.04)}$&$\mathit{  3.76\pm  0.02}$&$\mathit{ -0.02\pm  0.02}$&$\mathit{  (0.06\pm  0.04)}$\\
S0&110&$\mathbf{  3.70\pm  0.02}$&$\mathbf{ -0.13\pm  0.02}$&-24.00&$\mathbf{  0.14\pm  0.02}$&$\mathbf{  3.67\pm  0.02}$&$\mathbf{ -0.12\pm  0.02}$&$\mathbf{  0.18\pm  0.02}$\\
&&$\mathit{  3.67\pm  0.02}$&$\mathit{ -0.10\pm  0.03}$&&$\mathit{  0.14\pm  0.03}$&$\mathit{  3.64\pm  0.02}$&$\mathit{ -0.11\pm  0.03}$&$\mathit{  0.16\pm  0.03}$\\
Sa& 94&$\mathbf{  3.52\pm  0.04}$&$\mathbf{ -0.22\pm  0.04}$&-23.85&$\mathbf{  0.37\pm  0.03}$&$\mathbf{  3.35\pm  0.04}$&$\mathbf{ -0.20\pm  0.03}$&$\mathbf{  0.29\pm  0.03}$\\
&&$\mathit{  3.51\pm  0.04}$&$\mathit{ -0.22\pm  0.05}$&&$\mathit{  0.36\pm  0.04}$&$\mathit{  3.35\pm  0.04}$&$\mathit{ -0.20\pm  0.05}$&$\mathit{  0.28\pm  0.04}$\\
Sb&140&$\mathbf{  3.53\pm  0.04}$&$\mathbf{ -0.14\pm  0.04}$&-24.31&$\mathbf{  0.37\pm  0.03}$&$\mathbf{  3.18\pm  0.03}$&$\mathbf{ -0.17\pm  0.03}$&$\mathbf{  0.30\pm  0.02}$\\
&&$\mathit{  3.51\pm  0.04}$&$\mathit{ -0.12\pm  0.05}$&&$\mathit{  0.36\pm  0.03}$&$\mathit{  3.17\pm  0.03}$&$\mathit{ -0.14\pm  0.04}$&$\mathit{  0.29\pm  0.03}$\\
Sbc& 91&$\mathbf{  3.36\pm  0.04}$&$\mathbf{ -0.19\pm  0.04}$&-24.12&$\mathbf{  0.35\pm  0.03}$&$\mathbf{  3.14\pm  0.03}$&$\mathbf{ -0.24\pm  0.03}$&$\mathbf{  0.26\pm  0.03}$\\
&&$\mathit{  3.33\pm  0.04}$&$\mathit{ -0.12\pm  0.07}$&&$\mathit{  0.32\pm  0.04}$&$\mathit{  3.12\pm  0.03}$&$\mathit{ -0.18\pm  0.06}$&$\mathit{  0.25\pm  0.03}$\\
Sc&102&$\mathbf{  3.22\pm  0.04}$&$\mathbf{ -0.19\pm  0.04}$&-23.44&$\mathbf{  0.39\pm  0.03}$&$\mathbf{  2.87\pm  0.03}$&$\mathbf{ -0.23\pm  0.03}$&$\mathbf{  0.27\pm  0.03}$\\
&&$\mathit{  3.19\pm  0.04}$&$\mathit{ -0.17\pm  0.05}$&&$\mathit{  0.38\pm  0.03}$&$\mathit{  2.85\pm  0.03}$&$\mathit{ -0.21\pm  0.04}$&$\mathit{  0.25\pm  0.03}$\\
Sd& 95&$\mathbf{  2.98\pm  0.05}$&$\mathbf{ -0.08\pm  0.04}$&-22.11&$\mathbf{  0.46\pm  0.04}$&$\mathbf{  2.65\pm  0.04}$&$\mathbf{ -0.13\pm  0.03}$&$\mathbf{  0.37\pm  0.03}$\\
&&$\mathit{  2.95\pm  0.05}$&$\mathit{ -0.07\pm  0.05}$&&$\mathit{  0.46\pm  0.04}$&$\mathit{  2.64\pm  0.04}$&$\mathit{ -0.12\pm  0.04}$&$\mathit{  0.34\pm  0.03}$\\
Im& 20&$\mathbf{  2.55\pm  0.09}$&$\mathbf{ -0.07\pm  0.05}$&-21.60&$\mathbf{  0.34\pm  0.07}$&$\mathbf{  2.32\pm  0.08}$&$\mathbf{ -0.09\pm  0.04}$&$\mathbf{  0.27\pm  0.06}$\\
&&$\mathit{  2.54\pm  0.08}$&$\mathit{ -0.06\pm  0.07}$&&$\mathit{  0.31\pm  0.09}$&$\mathit{  2.32\pm  0.08}$&$\mathit{ -0.08\pm  0.06}$&$\mathit{  0.24\pm  0.11}$\\
\end{tabular}						   	 
\caption{$(B-H)_{\mathrm{T}}$ and $(B-H)^0_{\mathrm{T}}$ per type as a function of $X_{i1}=H_{\mathrm{abs}}$.}
\label{tabCMtype}
\end{center}
\end{table*}
We have plotted the $(B-H)_{\mathrm{T}}$-$H_{\mathrm{abs}}$ 
relation as a function
of type in Fig.~\ref{figCMtypeNIR}. We also show in
Fig.~\ref{figCMtypeNIRcorr} the
$(B-H)^0_{\mathrm{T}}$-$H_{\mathrm{abs}}$ 
relations where $(B-H)^0_{\mathrm{T}}$ is the color corrected to
face-on using the $\beta_1[\logd(\Rvc)]$ determined
in~\ref{corrinclin}. We assume that the extinction is negligible in
the NIR and do not correct $H_{\mathrm{abs}}$. No correction is also
applied to elliptical colors because their color-$\Rvc$ relation is not
due to the extinction.
The values of the estimators are given in Table~\ref{tabCMtype}.
We note that the intrinsic scatter in the corrected colors 
is significantly reduced. 
A more picturesque comparison of the relations for the different types
is plotted on Fig.~\ref{segments}. Though the agreement between the estimators
is not as good as previously and the uncertainties are large, the
slope obviously depends on the type. We tentatively summarize the
following results:
\paragraph{Star-forming galaxies:}
the slope is much flatter for 
Sd--Im than for Sa--Sc galaxies. Mobasher et al. (\cite{Mobasher}) 
probably did not detect this because the few Sd/Im galaxies
in their late-spirals sample were lost among Sbc and Sc galaxies.
At a given NIR absolute magnitude, the mean colors of the
types are different, indicating that the CM relation is not simply a 
type-mass relation: the NIR intrinsic luminosity must
be completed by the type to characterize the colors (and the star 
formation history) of star-forming galaxies.
A similar conclusion was drawn by Gavazzi (\cite{Gavazzi93})
but was later challenged by Gavazzi et al. (\cite{Gavazzi96b}).
\paragraph{E/S0:}
the color-magnitude relation of early-type galaxies is usually
explained by the galactic winds produced by super-novae 
in a starbursting environment, which expel the gas and quench the
star formation (Matthews \& Baker \cite{MB}).
Massive (and NIR bright) galaxies 
having a deeper gravitational well, they are able to retain 
the gas longer and to prolong the star-forming phase. This results
in a higher mean stellar metallicity and thus in redder colors
(Faber \cite{Faber}).
Besides its interest to constrain the models of galaxy formation,
the CM relation has also been proposed as a distance indicator.

We find a different slope for ellipticals and lenticulars.
The slope of S0 seems 
closer to that of early and intermediate spirals than to that of
ellipticals, which may give some clues on their evolution.

The MCES estimation does not detect any significant slope for
ellipticals. Because of the small intrinsic scatter, it is maybe not
the best estimator\footnote{Note that the slopes obtained for the 
{\em effective} 
color vs. absolute magnitude relation with the two estimators 
($\mathbf{-0.03}/\mathit{-0.02}$)
are in good agreement, both between them and with the MCES slope 
of the total color-magnitude relation.}, but even 
the slope determined by ML is smaller than that of the
comparable $(V-K)[5h^{-1}\,\mathrm{kpc}]$-$V_{\mathrm{T}}$ relation 
established by
Bower et al. (\cite{Bower}) in the Virgo and Coma clusters. 
Our intrinsic scatter is also larger that theirs. A possible
explanation of this discrepancy is that our sample contains 
not only cluster galaxies but also field ellipticals, which may have a
different star formation history and increase the scatter (Larson et
al. \cite{Larson}; Kauffmann \& Charlot \cite{KC}; Baugh et
al. \cite{Baugh}; see yet
Bernardi et al. \cite{Bernardi} and Schade et al. \cite{Schade}).
However, we also note that
the relation established by Bower et al. is based on {\em aperture}
colors. Their aperture is comparable to the mean effective aperture 
of ellipticals in Virgo and samples only about half their luminosity
(cf. also Kodama et al. \cite{Kodama}).
Since the effective aperture of bright ellipticals is larger, only
their inner regions are observed in the Bower et al.'s aperture.
Because of the blue-outwards color gradient, their aperture color is redder
and the slope is higher than the one derived from total colors.
We therefore encourage the modellists of elliptical galaxies, either to
compare their relations to 
total observed colors, or to predict aperture colors.
\begin{figure*}
\resizebox{!}{22cm}{\rotatebox{0}{\includegraphics{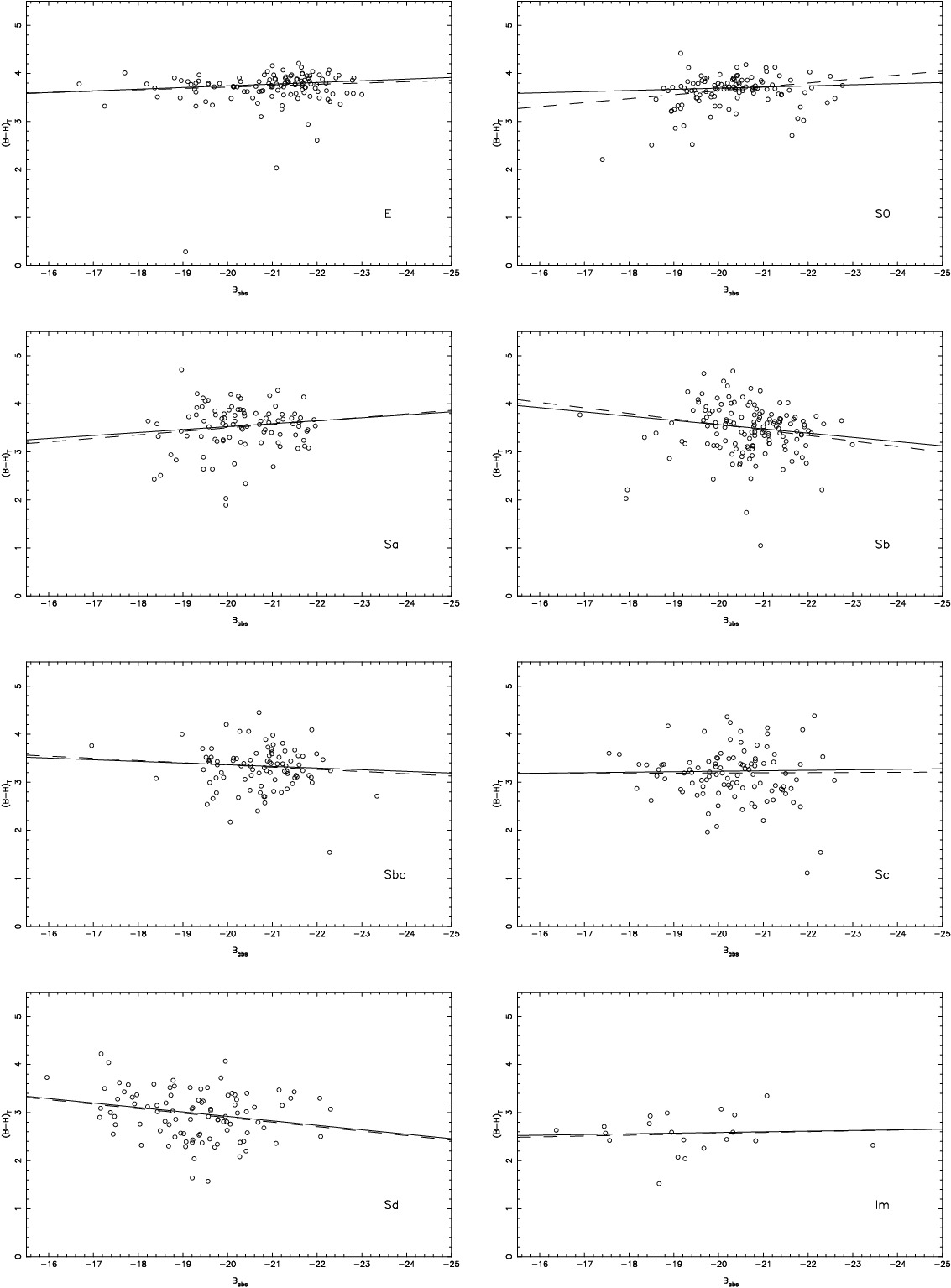}}}
\caption{$(B-H)_{\mathrm{T}}$-absolute $B$ magnitude
relations per type.}
\label{CMtypeB}
\end{figure*}

\begin{figure*}
\resizebox{!}{22cm}{\rotatebox{0}{\includegraphics{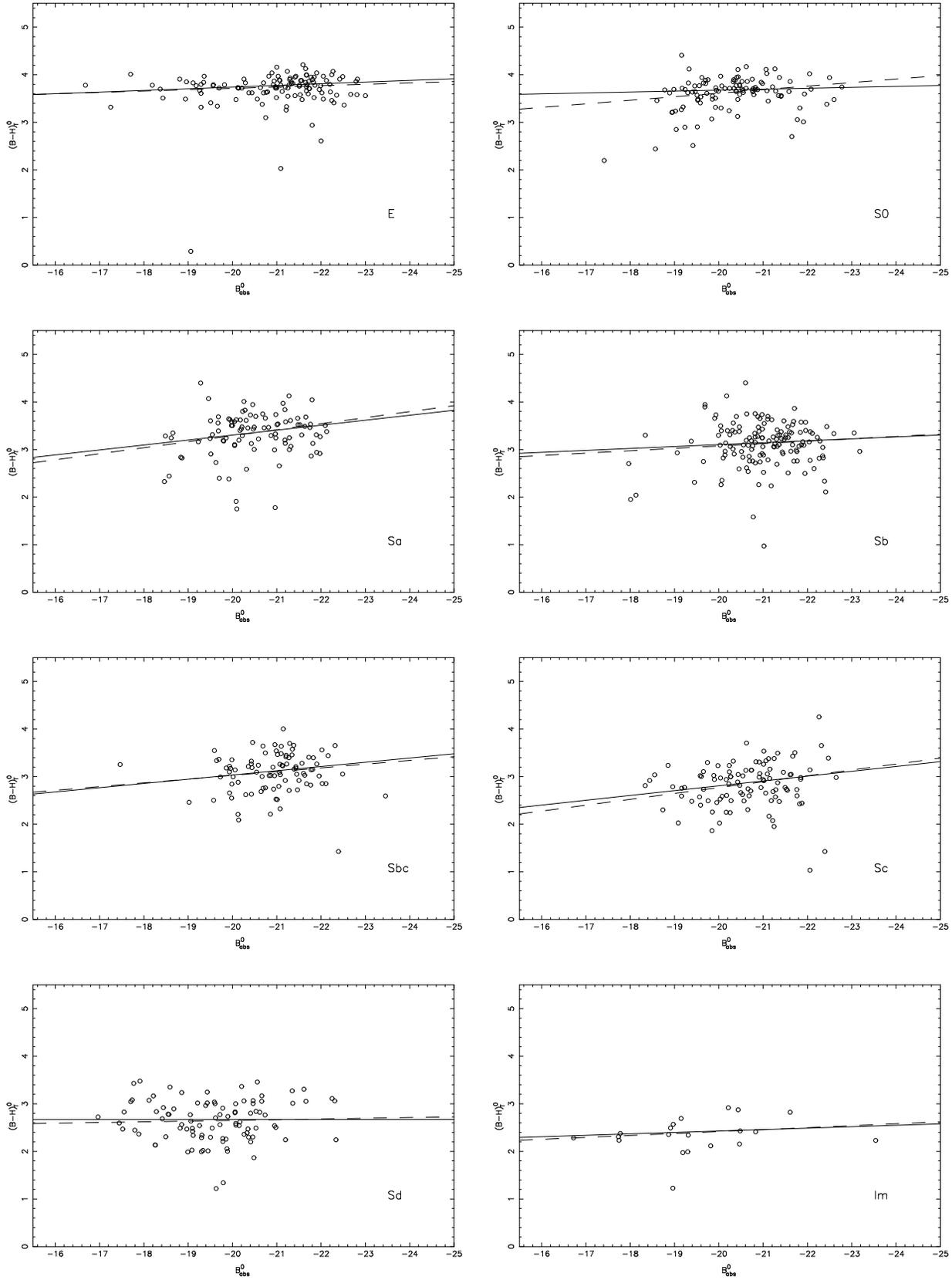}}}
\caption{$(B-H)^0_{\mathrm{T}}$-absolute $B^0$ magnitude
relations per type.}
\label{CMtypeBcorr}
\end{figure*}

The color-magnitude relations as a function of the $B$ absolute
magnitude are finally plotted for each type on figure~\ref{CMtypeB}.
The slopes are very different of those obtained when the
NIR is the reference band. For spiral galaxies especially,
they tend to have a positive sign, with brighter galaxies (in $B$)
being bluer! This behavior is typically expected if the 
relation is dominated by the extinction. 
If we assume that the extinction is negligible in the NIR, the
same correction applied to colors may be used to compute face-on
absolute magnitudes $B^0_{\mathrm{abs}}$.
The $(B-H)^0_{\mathrm{T}}$-$B^0_{\mathrm{abs}}$ relations per type show then
more reasonable slopes (Fig.~\ref{CMtypeBcorr}).
\begin{table*}
\begin{center}
\begin{tabular}{l|r|c|c|c|c|c|c}
type & $N$ &$\beta_0[(B-H)_{\mathrm{T}}]$&$\beta_1(H_{\mathrm{abs}})$&$\mu_1$&$\beta_2[\logd(\Rvc)]$&$\mu_2$&$\sigma$\\
\hline
E&115&$\mathbf{  3.80\pm  0.02}$&$\mathbf{ -0.05\pm  0.01}$&-24.97&$\mathbf{ -0.67\pm  0.19}$&  0.10&$\mathbf{  0.10\pm  0.02}$\\
&&$\mathit{  3.76\pm  0.02}$&$\mathit{ -0.02\pm  0.02}$&&$\mathit{ -0.83\pm  0.35}$&&$\mathit{  (0.04\pm  0.03)}$\\
S0&106&$\mathbf{  3.66\pm  0.02}$&$\mathbf{ -0.16\pm  0.02}$&-24.05&$\mathbf{  0.37\pm  0.16}$&  0.19&$\mathbf{  0.18\pm  0.02}$\\
&&$\mathit{  3.66\pm  0.02}$&$\mathit{ -0.12\pm  0.04}$&&$\mathit{  0.27\pm  0.17}$&&$\mathit{  0.14\pm  0.03}$\\
Sa& 94&$\mathbf{  3.47\pm  0.04}$&$\mathbf{ -0.23\pm  0.03}$&-23.85&$\mathbf{  1.18\pm  0.23}$&  0.17&$\mathbf{  0.30\pm  0.03}$\\
&&$\mathit{  3.49\pm  0.04}$&$\mathit{ -0.21\pm  0.04}$&&$\mathit{  1.21\pm  0.20}$&&$\mathit{  0.27\pm  0.04}$\\
Sb&140&$\mathbf{  3.47\pm  0.03}$&$\mathbf{ -0.17\pm  0.03}$&-24.31&$\mathbf{  0.97\pm  0.13}$&  0.25&$\mathbf{  0.30\pm  0.02}$\\
&&$\mathit{  3.45\pm  0.03}$&$\mathit{ -0.14\pm  0.04}$&&$\mathit{  0.99\pm  0.13}$&&$\mathit{  0.29\pm  0.02}$\\
Sbc& 91&$\mathbf{  3.30\pm  0.03}$&$\mathbf{ -0.28\pm  0.03}$&-24.12&$\mathbf{  1.22\pm  0.16}$&  0.25&$\mathbf{  0.23\pm  0.03}$\\
&&$\mathit{  3.28\pm  0.03}$&$\mathit{ -0.22\pm  0.06}$&&$\mathit{  1.19\pm  0.21}$&&$\mathit{  0.24\pm  0.03}$\\
Sc&102&$\mathbf{  3.13\pm  0.04}$&$\mathbf{ -0.24\pm  0.03}$&-23.44&$\mathbf{  1.10\pm  0.13}$&  0.29&$\mathbf{  0.27\pm  0.03}$\\
&&$\mathit{  3.11\pm  0.03}$&$\mathit{ -0.22\pm  0.04}$&&$\mathit{  1.17\pm  0.14}$&&$\mathit{  0.24\pm  0.03}$\\
Sd& 95&$\mathbf{  2.91\pm  0.04}$&$\mathbf{ -0.14\pm  0.03}$&-22.11&$\mathbf{  1.15\pm  0.16}$&  0.32&$\mathbf{  0.33\pm  0.03}$\\
&&$\mathit{  2.88\pm  0.04}$&$\mathit{ -0.15\pm  0.04}$&&$\mathit{  1.23\pm  0.16}$&&$\mathit{  0.33\pm  0.03}$\\
Im& 20&$\mathbf{  2.52\pm  0.07}$&$\mathbf{ -0.07\pm  0.04}$&-21.60&$\mathbf{  1.11\pm  0.29}$&  0.25&$\mathbf{  0.18\pm  0.05}$\\
&&$\mathit{  2.53\pm  0.06}$&$\mathit{ -0.05\pm  0.05}$&&$\mathit{  1.09\pm  0.29}$&&$\mathit{  (0.04\pm  0.06)}$\\
\end{tabular}
\caption{$(B-H)_{\mathrm{T}}$ color per type as a function of
$X_{i1}=H_{\mathrm{abs}}$ and $X_{i2}=\logd(\Rvc)$.}
\label{CMinclin}
\end{center}
\end{table*}
\subsection{The color vs. $\Rvc$ \& absolute magnitude relation}
We finally give in Table~\ref{CMinclin} the estimators for a
simultaneous fitting of the $(B-H)_{\mathrm{T}}$ colors as a function
of $\logd(\Rvc)$ and $H_{\mathrm{abs}}$ in each type.The overall
agreement with the slopes determined previously from the separate
fitting of the colors with respect to either $\Rvc$ or
$H_{\mathrm{abs}}$ is correct. The requirement that $H_{\mathrm{abs}}$
be known tends to select the brightest galaxies and may explain 
the small discrepancies in the slope as as a function of $\logd(\Rvc)$,
as this one depends probably also on the mass ($\sim
H_{\mathrm{abs}}$) of the galaxy.
The very small scatter obtained for Im galaxies indicates that there
are too few of them for the simultaneous analysis.
\section{Discussion}
\label{discussion}
NIR growth curves of the magnitude as a function of the
aperture have been built from the Catalog of Infrared Observations (Gezari et
al. \cite{Gezari}) and optical catalogs. Using these, we have been able
to compute total NIR apparent and absolute magnitudes, NIR and optical-to-NIR colors, and to
estimate their uncertainties for a large sample. A statistical
analysis of the colors as a function of type, inclination and
luminosity, using estimators taking into account the uncertainties
in the variables and
the intrinsic scatter in the colors, highlights the interest of the NIR and
notably of the comparison of the optical to the NIR.

Optical-to-NIR colors show a well defined sequence with type: the
mean irregular is 1.3 magnitude bluer in $B-H$ than the mean elliptical.
The intrinsic scatter is higher for star-forming galaxies ($\sigma\sim
0.4$) than for
ellipticals or lenticulars ($\sigma\sim 0.1$--$0.2$).

Because of the small extinction in the NIR, the optical-to-NIR colors
of spiral galaxies redden considerably with increasing inclination,
putting therefore constraints on the amount of the dust and the
respective distributions of dust and stars. S0 colors 
do not depend on the inclination. Rounder elliptical are redder than 
more elongated ones.

The color-absolute magnitude relation is much steeper and tighter
when the NIR is used as the reference band rather than the optical.
Examination of the color vs. $B$ absolute magnitude relation for each type
suggests that this is due to the extinction which dims and reddens 
the galaxies and counteracts the color-NIR relation.
A color-magnitude relation exists in each type, with brighter galaxies
in the NIR being redder. The slope is steeper for Sa--Sc than for 
early-type galaxies or Sd--Im. The relation we obtain for ellipticals is shallower
than in other studies. A likely explanation 
is that this is due to the effect of the color gradient
in small apertures, a lesser, i.e. redder, fraction 
of the galaxy being observed in large bright ellipticals 
than in small faint ones
-- a problem we do not have using our 
total (or we expect so) colors. The slope for S0 galaxies
is closer to that of spirals than to that of ellipticals.
The color at a given absolute magnitude becomes
bluer with increasing type, indicating that both the mass and the type
must be used to characterize the colors and to describe
the star formation history of galaxies.
Once corrected for the inclination and the magnitude effect, 
the intrinsic scatter in the colors of spiral types drops
significantly, although a few outliers with very blue colors typical of
starbursting galaxies remain.

However satisfying these results are, some problems still exist.
NIR observations are obtained in small apertures and
the extrapolation to total magnitudes and colors suffers from significant
uncertainties. Nevertheless, we do not believe that our results 
strongly depend on this extrapolation since similar slopes
are obtained when we use effective colors -- which are interpolated or
only slightly extrapolated -- instead of total colors.

The selection of galaxies is a more crucial problem.
The only criterion used here has been to take whatever was available
in the NIR.
For this reason, the sample, especially in $J$ and $K$, contains
few late-type galaxies but many cluster galaxies, Seyferts
and other active nuclei. 
We also lack of intrinsically faint galaxies. The 
color-magnitude relations we have determined should therefore
not be extrapolated at $H_{\mathrm{abs}}$ fainter than $\sim -20$.
Large samples as the Sloan Digital Sky Survey in the optical
or DENIS and 2MASS in the NIR should solve this problem.

The last important problem is the bias between the two estimators
we use, especially in the slope of the color-magnitude relation. 
The discrepancy tends to disappear however when we consider
effective colors which have smaller uncertainties. The MCES 
estimators seem to be closer to the true slopes than the ML,
but the latter provides a better estimate of the intrinsic 
scatter and has a lower variance.

Combining the NIR and the optical enlarges considerably our vision 
of the stellar populations and the dust content of galaxies. 
In our next paper, we will extend our analysis to the near 
ultraviolet-optical wavelength range. We will publish ``color''
energy distributions of galaxies from the near-UV to the NIR
that will be used as templates to constrain the
star formation history of galaxies of different types and masses.
A further step will be the extension of these techniques 
to the far-UV and the far-IR.
\begin{acknowledgements}

This research has made use of:
\begin{itemize}
\item
the Catalog of Infrared Observations operated at the NASA/Goddard
Space Flight Center (Gezari et al. \cite{Gezari}),
\item
the Hypercat catalog operated at the CRAL-Observatoire
de Lyon (Prugniel \& H\'eraudeau \cite{PH}),
\item
the Third Reference Catalog of Bright Galaxies 
(de Vaucouleurs G. et al. \cite{RC3}),
\item
the NASA/IPAC Extragalactic Database
(NED) which is operated by the Jet Propulsion Laboratory, 
California Institute of Technology, 
under contract with the National 
Aeronautics and Space Administration.
\end{itemize}

We also acknowledge:
\begin{itemize}
\item
the Statistical Consulting
Center for Astronomy operated at the Department of Statistics,
Penn State University, M.G. Akritas (Director),
\item
Numerical Recipes by Press et al. (\cite{NR}).
\end{itemize}

M. F. acknowledges support from the
National Research Council through the Resident Research Associateship Program.
\end{acknowledgements}
\appendix
\section{The optical growth curve}
\label{CCvis}
Let ${\cal S}_n(X)$ be defined by 
\[{\cal S}_n(X)=-2.5\logd\left[\frac{\int_0^{10^{(X-X_0)/n}}
x^{2n-1}\exp(-x)\,\mathrm{d}x}{\int_0^{\infty}x^{2n-1}
\exp(-x)\,\mathrm{d}x}\right]\]
where $X_0$ is such as ${\cal S}_n(0)=2.5\logd(2)=0.75$.

According to Prugniel \& H\'eraudeau (\cite{PH}), 
${\cal B}(X,\Tp)$ is defined by:
\[{\cal B}(X,\Tp)={\cal S}_n(X),\;n=(15-\Tp)/5\] 
if $\Tp<-5$ or $\Tp>10$
-- corresponding to a S\'ersic (\cite{Sersic}) luminosity profile 
$i(A)\propto\exp\left[-(A/A_0)^{1/n}\right]$ --,
and by \[{\cal B}(X,\Tp)=(2/3-\Tp/15){\cal S}_4(X)+(1/3+\Tp/15){\cal
S}_1(X)\] if $-5<\Tp<10$, i.e., interpolated between the de
Vaucouleurs profile and the exponential profile valid respectively for
giant ellipticals and pure-disk galaxies.
\section{Computation of NIR magnitudes and uncertainties}
\label{calculmag}
Let us first define ${\cal B}_i={\cal B}(X_i,\Tp)$,
\[S=\sum_{i=1}^n w_i,\] 
where $w_i$ is the weight attributed to the $i^{\mathrm{th}}$ point,
\[S_{\cal B}=\sum_{i=1}^n w_i{\cal B}_i,\]
\[S_{{\cal B}^2}=\sum_{i=1}^n w_i{\cal B}_i^2,\]
\[S_m=\sum_{i=1}^n w_i m_i,\]
\[S_{{\cal B}m}=\sum_{i=1}^n w_i{\cal B}_im_i,\]
and
\[\Delta=SS_{{\cal B}^2}-S_{\cal B}^2.\]

Two methods have been used, depending essentially on
the number of observations and their accuracy. 
In the first one,
the total NIR magnitude $m_{\mathrm{T}}$ is computed assuming $s=s_0$
by minimizing
\[\sum_{i=1}^{n}w_i\left(m_i-m_{\mathrm{T}}-s_0{\cal
B}_i\right)^2\] with respect to $m_{\mathrm{T}}$.
We obtain
\[m_{\mathrm{T}}^{(1)}=(S_m-s_0S_{\cal B})/S,\]
\[\sigma^2\left[m_{\mathrm{T}}^{(1)}\right]=
\left(\sigma_{s_0}^2 S_{\cal B}^2+\sum_{i=1}^n w_i^2\sigma_i^2\right)/S^2,\]
and the corresponding effective magnitude is 
\[m_{\mathrm{e}}^{(1)}=m_{\mathrm{T}}+2.5\logd(2)s_0,\]
\[\sigma^2\left[m_{\mathrm{e}}^{(1)}\right]=
\left\lbrace\sigma_{s_0}^2 \left[2.5\logd(2)S-S_{\cal B}\right]^2
+\sum_{i=1}^n w_i^2\sigma_i^2
\right\rbrace/S^2.\] 
We assume here that, for want of constraint on $s$,
the uncertainty on $s$ ($\sigma(s)$) is equal to the intrinsic
scatter $\sigma_{s_0}$ for this type.

In the second method, $s$ is considered as a free parameter
and $m_{\mathrm{T}}$ is computed by minimizing 
\[\sum_{i=1}^{n}w_i\left(m_i-m_{\mathrm{T}}-s{\cal
B}_i\right)^2\]
with respect to $m_{\mathrm{T}}$ and $s$.
We then get
\[m_{\mathrm{T}}^{(2)}=(S_{{\cal B}^2}S_m-S_{\cal B}S_{{\cal B}m})/\Delta,\]
\[\sigma^2\left[m_{\mathrm{T}}^{(2)}\right]=\left[\sum_{i=1}^n w_i^2\sigma_i^2\left(S_{{\cal B}^2}-{\cal
B}_i S_{\cal B}\right)^2\right]/\Delta^2,\]
\[s^{(2)}=(S S_{{\cal B}m}-S_{\cal B}S_m)/\Delta,\]
\[\sigma^2\left[s^{(2)}\right]=\left[\sum_{i=1}^n w_i^2\sigma_i^2\left(S_{\cal B}-{\cal
B}_i S\right)^2\right]/\Delta^2,\]
\[m_{\mathrm{e}}^{(2)}=m_{\mathrm{T}}+2.5\logd(2)s,\]
and
\begin{eqnarray*}
\sigma^2\left[m_{\mathrm{e}}^{(2)}\right]=\left\lbrace
\displaystyle{\sum_{i=1}^n}w_i^2\sigma_i^2\right.
\!\!&\!\!\left[\right.\!\!&\!\! S_{{\cal B}^2}-{\cal B}_i S_{\cal B}\\
\!\!&\!\!\mbox{}+\!\!&\!\!\left.\left.2.5\logd(2)\left(S{\cal
B}_i-S_{\cal B}\right)\right] ^2\rule{-2pt}{17pt}\right\rbrace
/\Delta^2.
\end{eqnarray*}
Note that we have made the questionable assumption
that the errors in the aperture magnitudes of a given galaxy
are independent.

We are however not primarily interested in the best fitting growth
curve, 
which would be determined assuming $w_i=1/\sigma_i^2$,
but rather in the best estimate of the asymptotic magnitude.
When there is only one point, the uncertainty on 
$m_{\mathrm{T}}$ is $\sigma(m_{\mathrm{T}})=(\sigma_1^2+{\cal B}_1^2
\sigma(s)^2)^{1/2}$, i.e., the uncertainty on $m_{\mathrm{T}}$
decreases with the size of the aperture. It is therefore reasonable
to adopt $w_i=1/(\sigma_i^2+{\cal B}_i^2\sigma(s)^2)$ for all the
points to give more weight to the large apertures.

To deal with outliers, we apply an iterative procedure. We initially
assume $\sigma(s)=\sigma_{s_0}$ for the slope and $\sigma_i=\sigma_{m_0}$ for all the points. 
At each step, we fit the growth curve to the data, compute $\sigma(s)$
from above equations,
and estimate new values of the uncertainties by
$\sigma_i=\max(\sigma_{m_0},|m_i-m_{\mathrm{T}}-s{\cal B}_i|)$.
This procedure is a compromise between our {\em a priori} uncertainty 
$\sigma_{m_0}$ and the {\em a posteriori} estimate $[\sum_{i=1}^n(m_i-s{\cal
B}_i)^2/ (n-k)]^{1/2}$, where $k=1$ or 2 is the number of fitted 
parameters (and also the index of the method!).
It has the advantage of reducing the weight 
of outliers and takes the scatter around the growth curve
automatically into account.
Practically, it improves the fit and gives a reasonable
estimate of the uncertainties.
The convergence is usually achieved in a few iterations. 

The second method provides a better fit 
than the first one but is less secure because the slope of the growth
curve is free.
It is used only when following conditions are fulfilled simultaneously:
\begin{enumerate}
\item
$n\ge 3$, which allows to check the validity of the fit,
\item
$\sigma\left[m_{\mathrm{T}}^{(2)}\right]<\sigma\left[m_{\mathrm{T}}^{(1)}\right],$
\item
and
\begin{eqnarray*}
\mathrm{e}^{-\left[s^{(2)}-s_0\right]^2/\left[2\sigma_{s_0}^2\right]}
&\!\!&
\prod_{i=1}^n 
\mathrm{e}^{-\left[m_i-m^{(2)}_{\mathrm{T}}-s^{(2)}
{\cal B}_i\right]^2/\left[2\sigma_{m_0}^2\right]}
\\
<&\!\!&
\prod_{i=1}^n 
\mathrm{e}^{-\left[m_i-m^{(1)}_{\mathrm{T}}-s^{(1)}{\cal
B}_i\right]^2/\left[2\sigma_{m_0}^2\right]}
,
\end{eqnarray*}
i.e., the fit provided by (2) is better enough than (1) to justify
the deviation of $s$ relatively to $s_0$.
\end{enumerate}

\end{document}